\documentclass[ aip, amsmath,amssymb, reprint,floatfix]{revtex4-1}
\usepackage{graphicx}
\usepackage{color}
\usepackage{amsmath}
\usepackage{braket}
\usepackage{enumerate}
\usepackage{hyperref}

\usepackage[caption=false]{subfig}
\usepackage{floatrow}
\floatstyle{plaintop}
\restylefloat{table}
\usepackage{subfig}
\usepackage[dvipsnames]{xcolor}
\usepackage[normalem]{ulem}

\newcommand{\Diosi}{Di\'{o}si}

\newcommand{\sigmaz}{\ensuremath{\sigma_{\mathrm{\mathrm{z}}}}}

\newcommand{\omegaq}{\ensuremath{\omega_{\mathrm{\mathrm{q}}}}}
\newcommand{\omegaqo}{\ensuremath{\omegaq^0}}

\newcommand{\EJ}{\ensuremath{E_\mathrm{J}}}
\newcommand{\EJo}{\ensuremath{E_\mathrm{J}^0}}
\newcommand{\EC}{\ensuremath{E_\mathrm{C}}}

\newcommand{\TTs}{\ensuremath{T_2^*}}
\newcommand{\gammaDP}{\ensuremath{\gamma_\mathrm{DP}}}

\newcommand{\omegam}{\ensuremath{\Omega}}			

\newcommand{\XZP}{\ensuremath{X_\mathrm{ZP}}}

\newcommand{\Teff}{\ensuremath{T_\mathrm{eff}}}

\newcommand{\Xmax}{\ensuremath{X_\mathrm{max}}}

\newcommand{\kB}{\ensuremath{k_\mathrm{B}} }

\newcommand{\Bpar}{\ensuremath{B_{||}} }

\newcommand{\ad}{\ensuremath{a^\dagger}}

\newcommand{\mA}{\ensuremath{m_\mathrm{a}} }
\newcommand{\tDP}{\ensuremath{t_\mathrm{DP}} }
\newcommand{\EG}{\ensuremath{E_\mathrm{G}} }		

\DeclareMathOperator{\Tr}{Tr}

\begin{document}
	
\title{Can the displacemon device test objective collapse models?}

\author{Lydia~A.~Kanari-Naish}
\email{l.kanari-naish18@imperial.ac.uk}
\affiliation{QOLS, Blackett Laboratory, Imperial College London, London SW7 2BW, United Kingdom}

\author{Jack~Clarke}
\email{jack.clarke@imperial.ac.uk}
\affiliation{QOLS, Blackett Laboratory, Imperial College London, London SW7 2BW, United Kingdom}

%
\author{Michael~R.~Vanner}
\email{m.vanner@imperial.ac.uk}
\affiliation{QOLS, Blackett Laboratory, Imperial College London, London SW7 2BW, United Kingdom}
%
%
\author{Edward~A.~Laird}
\email{e.a.laird@lancaster.ac.uk}
\affiliation{Department of Physics, Lancaster University, Lancaster, LA1 4YB, United Kingdom}

\begin{abstract}
Testing the limits of the applicability of quantum mechanics will deepen our understanding of the universe and may shed light on the interplay between quantum mechanics and gravity. At present there is a wide range of approaches for such macroscopic tests spanning from matter-wave interferometry of large molecules to precision measurements of heating rates in the motion of micro-scale cantilevers.
The ``displacemon" is a proposed electromechanical device consisting of a mechanical resonator flux-coupled to a superconducting qubit enabling generation and readout of mechanical quantum states.
%
In the original proposal, the mechanical resonator was a carbon nanotube, containing $10^6$ nucleons.
Here, in order to probe quantum mechanics at a more macroscopic scale,
we propose using an aluminium mechanical resonator on two larger mass scales, one inspired by the Marshall-Simon-Penrose-Bouwmeester moving-mirror proposal, and one set by the Planck mass. For such a device, we examine the experimental requirements needed to perform a more macroscopic quantum test and thus feasibly detect the decoherence effects predicted by two objective collapse models: \Diosi-Penrose and continuous spontaneous localization.
Our protocol for testing these two theories takes advantage of the displacemon architecture to create non-Gaussian mechanical states out of equilibrium with their
environment and then analyzing the measurement statistics of a superconducting qubit.
We find that with improvements to the fabrication and vibration sensitivities of these electromechanical devices, the displacemon device provides a new route to feasibly test decoherence mechanisms beyond standard quantum theory.
\end{abstract}

\date{\today{}}
\maketitle

\section{Introduction}
The quantum measurement problem and the quantum-to-classical interface has been much debated since the inception of quantum mechanics. This debate has sparked numerous questions such as: ``Is there a limit to the extent that the Schr{\"o}dinger equation may be applied?" and: ``What are the mechanisms that cause a coherent quantum superposition to lose its interference features and result in classical behaviour?". This debate and these particular questions explore the two key dynamical processes in quantum mechanics, namely the linear deterministic Schr{\"o}dinger evolution and the nonlinear stochastic wavefunction collapse owing to a measurement. 
Several different physical interpretations of these two processes have become popular among various groups within the physics and philosophy communities. However it is often very difficult or even impossible to experimentally distinguish between these differing views of the world.

One resolution to the measurement problem and the quantum-to-classical interface is provided by objective collapse theories.
These propose physical mechanisms, not encompassed by standard Schr{\"o}dinger evolution, which cause wavefunctions to collapse without the need for collapse by a measurement. 

Following early progress in modifying the Schr{\"o}dinger equation~\cite{Pearle1976,Pearle1979,Pearle1984,Gisin1984}, spontaneous localization models were studied by Ghirardi, Rimini, and Weber (GRW)~\cite{Ghirardi1986}. GRW theory satisfies the requirements of a physical collapse model with the key ingredient being that an $N$-particle wavefunction $\psi_{N}(x_1,\ldots,x_N)$ experiences a spontaneous Gaussian localization with the effect of premultiplying it by an operator of the form $\exp(-(X-x_k)^2/(2r^2))$, where $r$ is the localisation length and $X$ is the position operator. These Gaussian jumps happen at randomly distributed times with a mean rate $\lambda$ which is chosen to preserve unitary time evolution for microscopic systems, but importantly accounts for classical features emerging as the number of particles in an object is increased~\cite{Schlosshauer2007}. 
A further refinement of this type of collapse theory combines the ideas of stochastic dynamical reduction together with GRW to obtain continuous spontaneous localization (CSL)~\cite{Ghirardi1990}, which is now a widely studied model. In contrast to GRW theory, CSL assumes the localization operations happen continuously in time. Like GRW, CSL behaviour is characterised by two phenomenological parameters, the collapse rate $\lambda_{\mathrm{CSL}}$ and the localization length $r_{\mathrm{CSL}}$. 
The goal of many experimental tests of CSL is to impose bounds on these constants. To this end, $r_{\mathrm{CSL}}=100$~nm and $\lambda_{\mathrm{CSL}}=10^{-8\pm 2}$ have been identified as an interesting part of parameter space to explore \cite{Fu1997,Adler2007,AdlerRamazanoglu2007,Carlesso2019}.

Another contender to reconcile the issues presented by the measurement problem, offered by \Diosi \cite{Diosi1984,Diosi1987} and later rederived through a different line of reasoning by Penrose~\cite{Penrose1996}, is now referred to as \Diosi-Penrose (DP) collapse. This gravity-induced collapse model is argued by Penrose to arise from the incompatibility of the theories of general relativity and quantum mechanics.
The essence of the theory is that for a massive particle in a superposition of two well-localised states, each state has an associated space-time geometry leading to a conflict with Einstein’s theory of general covariance~\cite{Howl2019}. This superposition of differing gravitational fields leads to an instability which induces wave-function collapse after a characteristic lifetime $t_{\mathrm{DP}}$. To quantify the difference between two space-times, Penrose has suggested using the gravitational self-energy from classical physics $E_{\mathrm{G}}$, and relating this to $t_{\mathrm{DP}}$ via $t_{\mathrm{DP}}=\hbar/E_{\mathrm{G}}$. 

Tests of objective collapse models aim to detect deviations from standard quantum theory. Non-interferometric tests focus on sensing the additional sources of decoherence which are predicted by collapse models, usually in the form of heating. Searching for this excess noise has led to experimental realisations including test masses on ultracold cantilevers~\cite{Vinante2020}, underground tests~\cite{Donadi2021}, and entangling macroscopic diamonds~\cite{lee2011entangling,Belli2016}. Meanwhile, a host of non-interferometric experiments have been proposed using a range of platforms including optomechanics~\cite{bahrami2014proposal}, superconducting devices~\cite{Wal2000}, ultracold atoms~\cite{laloe2014heating}, and levitated nanospheres~\cite{Li2016}. Notably in optomechanics, building on the Bose-Jacobs-Knight proposal~\cite{Bose1999}, the Marshall-Simon-Penrose-Bouwmeester mirror-superposition scheme proposes how to study spatial superposition states involving $10^{14}$ atoms~\cite{Marshall2003}.
On the other hand, interferometric tests strive to create large superpositions with increasingly massive objects in order to probe the quantum-to-classical transition~\cite{Bassi2013}; these experiments include atom interferometry~\cite{kovachy2015quantum}, superpositions of large molecules~\cite{Eibenberger2013,Fein2019}, interferometry of free-falling nano-particles~\cite{Bateman2014}, and also proposals with ultracold atomic systems~\cite{bilardello2017collapse}. Meanwhile, electromechanical devices have been used in several experiments that, while not testing collapse models directly, nevertheless probe fundamental aspects of quantum behaviour, such as coherent exchange between a mechanical resonator and a qubit~\cite{OConnell2010}, and entanglement between a resonator and a microwave field~\cite{Palomaki2013} or another resonator~\cite{Ockeloen-Korppi2018}. Moreover, there is growing evidence that electromechanical devices are a viable route for tests of collapse, particularly with mechanical oscillators coupled to superconducting circuits~\cite{gely2021superconducting}.

In this work we provide an experimental scheme to probe collapse models using the displacemon platform~\cite{Khosla2018}. This device consists of a vibrating nanobeam coupled to a superconducting qubit such that operations performed on the qubit allow us to generate non-Gaussian states of the centre-of-mass of the mechanical system. We begin with a brief overview of this device and the qubit operations required to produce mechanical superpositions. A scheme is then presented that allows us to monitor the decoherence of these large-scale mechanical superpositions via qubit measurement and thereby test for sources of decoherence that could be attributed to DP or CSL collapse theories. We carefully examine the feasibility of our protocol given current and near-term experimental parameters and find that with technical improvements in isolating the device from vibrational noise, the displacemon device provides an experimentally viable route for testing DP and CSL models of objective collapse beyond existing bounds.

\section{Grating operations with the displacemon}
\subsection{Relationship between the displacemon and a conventional matter-wave interferometer}

A matter-wave interferometer works by passing the test particle through a series of diffraction gratings and then measuring its position.
A well-known example is the electron double-slit interferometer.
Quantum superposition manifests itself as a series of spatial fringes, which arise from interference between different paths a particle could take through the gratings.
This kind of matter-wave interferometry, now realised using increasingly massive test particles in sophisticated interferometers, allows the most stringent interferometry-based exclusions of spontaneous collapse theories~\cite{Arndt2014}.
The largest mass tested so far is 27,000~amu, surviving in a superposition for 7~ms~\cite{Fein2019}.
However, this mass is still too small to test gravitational decoherence and some theories of spontaneous collapse.

Progress in this area continues to be rapid, but to explore the gravitational-decohererence scale using conventional matter-wave interferometers, two major challenges need to be addressed.
First, as test masses become larger, it becomes increasingly difficult to create suitable diffraction gratings.
Existing state-of-the-art interferometers use finely patterned membranes in the first and final step.
However, the test particles interact with these gratings by van der Waals interaction, which becomes stronger as the particle diameter increases and introduces unwanted smearing of the wavefunction~\cite{Brezger2002}.
Future proposed interferometers will use all-optical gratings~\cite{Bateman2014}, but here the large mismatch between the optical wavelength and the de Broglie wavelength of a massive test particle imposes other stringent requirements on the interferometer.
The second challenge is the requirement for the particle to evolve in free-fall, so that the wavefunction has time to spread out before the final measurement.

The displacemon~\cite{Khosla2018} provides an approach to circumvent both these challenges by using as its test particle a doubly-clamped cantilever coupled to a superconducting qubit.
The key idea is to create a synthetic diffraction grating, realised through a series of operations on the qubit.
Figure~\ref{fig:displacemon}(a) summarises the principle, and how it relates to a conventional matter-wave interferometer.
In a conventional interferometer, a test particle moving along the axis of the interferometer encounters a series of gratings, each of which either blocks it or allows it to pass.
(In an optical grating, the particle may be ejected from the apparatus rather than blocked, but the effect is the same.)
Each grating effectively performs a projective measurement of the particle's position, in a basis that contains the states \{``blocked", ``passed"\}
\footnote{Of course, the state {``blocked"} will rapidly be further projected to the particular location at which the particle has struck the grating, but since trajectories that evolve via this state do not contribute to the final interference pattern, this process is not important for our purposes.}.
The process of passing through the gratings and then registering the particle's position is thus equivalent to a series of projective position measurements, with the final measurement outcome recorded only if all previous measurements have projected to ``passed" (i.e. the particle makes it through the interferometer unimpeded).
Figure~\ref{fig:displacemon}(a-b) illustrates this equivalence.

\subsection{The displacemon protocol}
\label{sec:displacemonprotocol}
\begin{figure}
	\centering
	\includegraphics{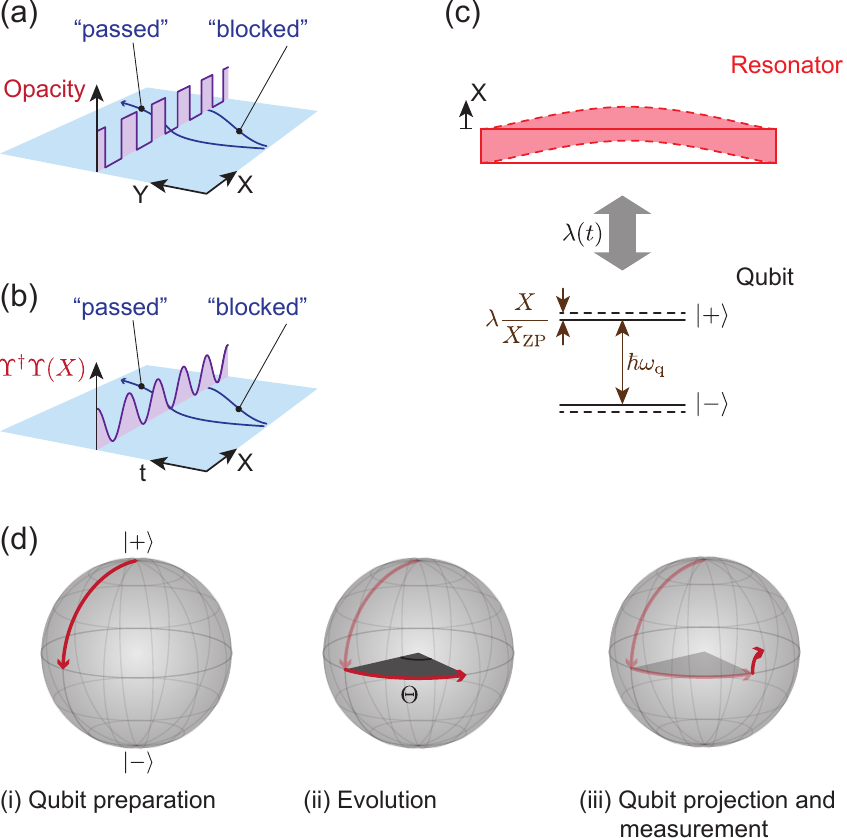}
	\caption{Implementing an effective diffraction grating.
		(a)~A transmission grating as in a conventional matter-wave interferometer.
		Particle trajectories in the $(X,Y)$ plane pass or are blocked depending whether they align with a slit or with the opaque region between slits.
		The encounter with the grating is a projective measurement, initially into the two states ``passed" and ``blocked".
		(b)~An effective grating realised by a projective measurement.
		Selecting based on the measurement filter $\Upsilon^\dagger\Upsilon(X)$ (see text) passes only those trajectories of the particle that align with the ``slits'' in the effective grating.
		(c)~Schematic of the coupled resonator-qubit device used to realise this operation.
		The resonator's displacement $X$ shifts the energy levels of the qubit with a strength given by the coupling rate $\lambda$.
		(d)~The grating protocol, which realises this measurement filter, illustrated via rotations on the Bloch sphere in the qubit's rotating frame.
		(i)~The qubit is prepared in a superposition state.
		(ii)~It precesses around the equator.
		At the end of this evolution it has precessed by an angle $\Theta$ which is proportional to the instantaneous displacement.
		(iii)~After a further rotation, the qubit state is measured.
		If the displacement is such as to induce an integer number of rotations, the qubit is likely to be found in state $\ket{+}$; if the displacement induces a half-integer number of rotations, the qubit is likely to be found as $\ket{-}$.
		This creates the periodic effective grating shown in (b).
		\label{fig:displacemon}
	}
\end{figure}

The Hamiltonian of the flux coupled qubit-resonator system is
\begin{equation}
	H = \hbar\frac{\omegaq}{2} \sigmaz + \hbar\omegam \ad a +\hbar \lambda(t) (a + \ad) \sigmaz~,
	\label{eq:Hamiltonian}
\end{equation}
where $\omegaq$ is the bare qubit frequency, $\sigmaz$ is the Pauli operator operating in the qubit basis with eigenvectors $\ket{+}$ and $\ket{-}$, and $\lambda$ is the qubit-mechanical coupling parameter, which may depend on time $t$.
The mechanical resonator has creation (annihilation) operator $a~(\ad)$ and mechanical frequency $\omegam/2\pi$.
The displacement is $X \equiv \XZP(a+\ad)$, where $\XZP \equiv \sqrt{\hbar/2m\Omega}$.
The interaction between the mechanical oscillator and the qubit shifts the energy splitting of the qubit by an amount proportional to the mechanical displacement (Fig.~\ref{fig:displacemon}(c)).

In the simplest implementation, the required projective measurement is realised as follows~(Fig.~\ref{fig:displacemon}(d)):
\begin{enumerate}[i.]
\item
The qubit is prepared in an eigenstate $(\ket{+} + \ket{-})/\sqrt{2}$.
\item
The system evolves under the Hamiltonian~\eqref{eq:Hamiltonian} for time $\tau_\mathrm{R}$.
In the qubit's rotating frame, it precesses at a rate proportional to the mechanical displacement $X$. (In the parameter regime considered here, the mechanical frequency is small enough that the particle can be assumed stationary during this step.
Ref.~\onlinecite{Khosla2018} explains how the protocol should be modified if this assumption is violated.)
At the end of this step, the accumulated qubit precession is
\begin{equation}
\Theta = \lambda \frac{X}{\XZP} \tau_\mathrm{R}~.
\end{equation}
\item
The qubit is subjected to a $\pi/2$ rotation about a horizontal axis and is projectively measured in the $\{\ket{+}, \ket{-}\}$ basis.
The probability to obtain the result $\ket{+}$ is
\begin{equation}
	p_+ = \cos^2 \Theta~.
\end{equation}
Conditioning on this outcome (i.e. analysing subsequent data only in those experimental runs for which this outcome was obtained) therefore projects the particle's displacement as shown in Fig.~\ref{fig:displacemon}(b), thus realising an effective grating. Conditioning on $\ket{-}$ results in a $\pi$ phase-shifted interference fringe.
\end{enumerate}

In a conventional interferometer, the particle encounters gratings that are located at different positions along its path (Fig.~\ref{fig:displacemon}(a)).
In the displacemon, the effective grating is turned on and off in time (Fig.~\ref{fig:displacemon}(b)).
In both cases, state preparation of the matter degree of freedom manifests itself as a diffraction pattern in the transverse coordinate $X$.

\section{The device}
We now explain the proposed device in detail and obtain expressions for each of the quantities in the Hamiltonian~\eqref{eq:Hamiltonian}.

\begin{figure*}
	\centering
	\includegraphics{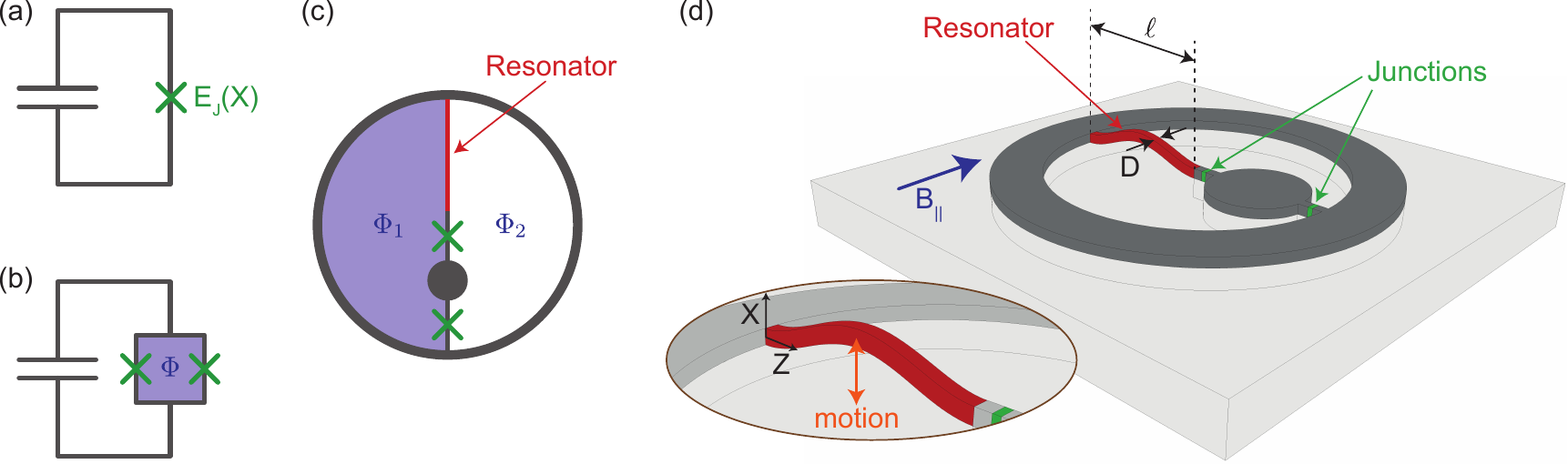}
	\caption{Evolution of the displacemon design
		(a) Principle of a mechanically coupled transmon qubit.
		The transmon is an $LC$ resonator in which the inductance is provided by a Josephson junction.
		If the Josephson energy $\EJ(X)$ depends on mechanical displacement, then so does the qubit frequency.
		(b) The mechanical dependence can be introduced by replacing the single Josephson junction by a SQUID.
		The Josephson energy now depends on the flux through the SQUID, which in turn depends on the displacement.
		(c) The displacemon design takes an additional step to decouple the qubit from homogeneous magnetic field fluctuations, which would otherwise perturb $\EJ$.
		That step is to incorporate two enclosed flux regions, so that the qubit energy depends on the flux difference between them, which is changed when the beam vibrates in a magnetic field,
		The two electrodes of the capacitor are the inner and outer circles.
		(d) Sketch of the displacemon analysed in this paper.
		Dark gray: fixed superconductor; Red: Vibrating superconducting beam; Green: Josephson junctions; Light gray: insulating substrate.
		Inset is a zoom-in on the moving resonator.
		\label{fig:device}
	}
\end{figure*}

\subsection{The resonator}
The resonator is modelled as a superconducting beam with length $\ell$, a square cross-section, and side length $D$ (Fig.~\ref{fig:device}(d)).
For simplicity, we assume that the beam has been fabricated without tension.
Its motion is described by the Euler-Bernoulli beam equation~\cite{Poot2012}:
\begin{equation}
\left(	\rho_{0} D^2 \frac{\partial^2}{\partial t^2} + \frac{E D^4}{12} \frac{\partial^4}{\partial Z^4}\right) u(Z,t)= 0~,
\end{equation}
where $u(Z,t)$ is the displacement profile of the beam as a function of time $t$ and coordinate along the beam $Z$, with $Z=0$ being the centre.
Here $\rho_{0}$ is the density of the beam and $E$ is its Young's modulus, and Poisson's ratio has been neglected.

To solve for the frequency of the fundamental displacement mode, we write
\begin{equation}
	u(Z,t) = u_0(Z) e^{i \Omega t}~,
\end{equation}
which means that $u_0(Z)$ must satisfy
\begin{equation}
	\frac{d^4}{d Z^4}  u_0(Z) = k^4 u_0(Z)~,
\end{equation}
where
\begin{equation}
	k \equiv \left(12\, \Omega^2 \frac{\rho_{0}}{ED^2}\right)^{1/4}.
	\label{eq:dispersionRelation}
\end{equation}
The solution, respecting the boundary conditions $u_0(\pm \ell/2)=0$
and normalising~\cite{Poot2012} so that $\int u_0^2(Z)\,dZ=1$, is
\begin{equation}
	u_0(Z) = \sqrt{2}\, \frac{\cos \frac{k \ell}{2} \cosh k Z - \cosh \frac{k \ell}{2} \cos k Z}{\sqrt{\cos^2 \frac{k \ell}{2} + \cosh^2 \frac{k \ell}{2}}}.
	\label{eq:u0}
\end{equation}
The additional boundary conditions $u_0'(\pm \ell/2)=0$
constrain $k$ to obey
\begin{equation}
	\cos \frac{k \ell}{2} \sinh \frac{k \ell}{2}  + \cosh \frac{k \ell}{2} \sin \frac{k \ell}{2}= 0,
\end{equation}
whose lowest solution is $k\ell\approx 4.730$.
Substituting this solution into Eq.~\eqref{eq:dispersionRelation} gives the frequency of the fundamental mode:
\begin{equation}
	\Omega \approx 6.47 \, \frac{D}{\ell^2} \, \sqrt{\frac{E}{\rho_{0}}}.
\end{equation}

An ideal square beam has two degenerate fundamental modes corresponding to orthogonal directions of motion.
However, only the out-of-plane mode couples strongly to the qubit.
\footnote{The reason is that only out-of-plane vibrations modulate the flux induced by $B_{||}$ (see Fig.~\ref{fig:device}(d)).
	In-plane vibrations couple to the perpendicular magnetic field, which is much weaker.}
Furthermore, fabrication imperfections will generally break the degeneracy of the two modes.
(If necessary, this can be enforced by designing the beam width to be different from its thickness.)
We therefore consider only the out-of-plane mode in the rest of this work.

\subsection{The qubit}
The qubit is based on the split concentric transmon design~\cite{Braumuller2016}.
Figure~\ref{fig:device} shows how this design relates to the conventional transmon~\cite{Koch2007, Krantz2019}.
A conventional transmon (Fig.~\ref{fig:device}(a)) consists of two superconducting islands coupled via a capacitance $C$ in parallel with a Josephson junction.
The Josephson junction acts as an effective non-linear inductor, meaning that the transmon is an anharmonic $LC$ resonator whose two lowest states form the qubit basis.
The qubit frequency is
\begin{equation}
	\omegaq=\frac{1}{\hbar} \sqrt{8\EC\EJ}~,
	\label{eq:qubitFrequency}
\end{equation}
where $\EC \equiv e^2/2C$ is the charging energy and $\EJ$ is the Josephson energy.
Clearly, if $\EJ$ depends on $X$, this realises the qubit-mechanical coupling envisaged in Section~\ref{sec:displacemonprotocol}.

One way to introduce such a coupling is to replace the single Josephson junction by a pair of junctions forming a superconducting quantum interference device (SQUID), making a `split transmon'\cite{DiCarlo2009} (Fig.~\ref{fig:displacemon}(b)).
Assuming each junction has Josephson energy $\EJo/2$, the Josephson energy of the combined junctions is
\begin{equation}
	\EJ = \EJo \left| \cos \frac{\pi\Phi}{\Phi_0}\right|,
\end{equation}
where $\Phi$ is the magnetic flux through the SQUID loop and $\Phi_0 \equiv h/2e$ is the superconducting flux quantum.
If a superconducting beam is introduced into one arm of the loop and the qubit is operated in a static magnetic field, then the displacement of the beam modulates $\Phi$ and therefore $\EJ$.
Electromechanical coupling between a vibrating beam and a transmon was recently demonstrated using this principle~\cite{Bera2021}.

A drawback of the circuit in Fig.~\ref{fig:device}(b) is that the qubit is sensitive to magnetic field noise, which modulates $\Phi$.
To mitigate this, the design is modified so that it contains two flux loops~(Fig.~\ref{fig:device}(c)), meaning that the two islands of the transmon are now the inner and outer circles.
Figure~\ref{fig:displacemon}(d) sketches a realisation of this circuit.
The effective Josephson energy is~\cite{Braumuller2016}
\begin{equation}
	\EJ=\EJo\left| \cos \frac{\pi\Delta\Phi}{2\Phi_0}\right|,
\end{equation}
where $\Delta\Phi \equiv \Phi_1 - \Phi_2$ is the difference in enclosed flux between the two loops.
From Eq.~\eqref{eq:qubitFrequency}, the qubit frequency is therefore
\begin{equation}
	\omegaq = \omegaqo \sqrt{\left| \cos \frac{\pi\Delta\Phi}{2\Phi_0}\right|},
	\label{eq:qubitFrequency2}
\end{equation}
where
\begin{equation}
	\omegaqo	=	\frac{1}{\hbar} \sqrt{8 \EJo \EC}
\end{equation}
is the qubit frequency when $\Delta \Phi = 0$.
The flux difference $\Delta \Phi(t)$, and therefore the qubit frequency, can be tuned using a small quasi-static perpendicular field gradient generated by an on-chip bias coil.

Such a qubit can be initialised, controlled, and measured using standard techniques of circuit quantum electrodynamics (cQED)~\cite{Krantz2019}.
In summary, the simplest initialisation is by thermal relaxation to the ground state $\ket{+}$.
Coherent control is achieved by applying microwave bursts resonant with $\omegaq$.
Projective readout in the $\sigmaz$ basis is achieved by coupling the qubit to an off-resonant cavity and measuring the state-dependent change to the cavity's resonant frequency.

\subsection{Qubit-resonator coupling}
The coupling described by Eq.~\eqref{eq:Hamiltonian} arises if the qubit frequency given by Eq.~\eqref{eq:qubitFrequency2} depends on the beam's displacement.
This occurs in the presence of an in-plane magnetic field $B_{||}$, applied orthogonal to the beam axis (Fig.~\ref{fig:displacemon}(d)).
In the presence of such a field, a displacement $X$ in the fundamental mode changes the flux difference by an amount
\begin{align}
	\delta \Phi &= \frac{d(\Delta \Phi))}{dX} X		\label{eq:PhiShift}\\
				&= \beta_0 {B_{||}} \ell X~,
\end{align}
where 
\begin{align}
	\beta_0 	&\equiv	\int_{-\ell/2}^{\ell/2} u_0(Z) \, dZ	\\
				&\approx		0.831
\end{align}
is a geometrical coupling factor that relates the displacement of the beam to the area it sweeps out.

Applying Eq.~\eqref{eq:PhiShift} to Eq.~\eqref{eq:qubitFrequency2} gives
\begin{align}
	\frac{d\omegaq}{dX} &= \frac{d\omegaq}{d(\Delta \Phi)} \, \frac{d(\Delta \Phi)}{dX}	\\
						&= -\omegaqo \, \frac{\pi \beta_0}{2} \,
										\frac{ \ell \Bpar}{\Phi_0} \, 
										\frac{\sin \frac{\pi \Delta \Phi}{2\Phi_0}}{\sqrt{\left|\cos \frac{\pi \Delta \Phi}{2\Phi_0}\right|}}~.
\end{align}
This allows us to calculate the electromechanical coupling rate in Eq.~\eqref{eq:Hamiltonian}:
\begin{align}
	\lambda(t) 	&\equiv \frac{\XZP}{2} \, \frac{d \omegaq}{dX}	\\
				&=		-\omegaqo \, \frac{\pi \beta_0}{4} \, \frac{\XZP\ell \Bpar}{\Phi_0} \, 
				\frac{\sin \frac{\pi \Delta \Phi(t)}{2\Phi_0}}{\sqrt{\left|\cos \frac{\pi \Delta \Phi(t)}{2\Phi_0}\right|}} \, \mathrm{sgn}\left({\cos\tfrac{\pi \Delta \Phi(t)}{2\Phi_0}}\right)\\
				&=      -\omegaq \frac{\pi \beta_0}{4} \, \frac{\XZP\ell \Bpar}{\Phi_0} \
				        \tan \frac{\pi \Delta \Phi(t)}{2\Phi_0}~,
				        \label{eq:lambdaValue}
\end{align}
where the time dependence arises through the tunable flux offset $\Delta \Phi (t)$.

In principle, Eq.~\eqref{eq:lambdaValue} allows arbitrarily large $\lambda(t)$.
In practice, the inevitable asymmetry between the two Josephson junctions has the effect of limiting the useful range of $\Delta \Phi$, meaning that the  $\tan \frac{\pi \Delta \Phi(t)}{2\Phi_0}$ factor is likely to be of order unity.

\subsection{Physical parameters}
\label{sec:PhysicalParameters}

\begin{table*}
\begin{center}
\caption{Parameters of the devices considered here, both suspended aluminium beams.
Device A is designed to have the same mass as the vibrating mirror proposed in Ref.~\onlinecite{Marshall2003}.
Device B is designed to have the Planck mass.
For both devices, the magnetic field is assumed to be $B_{||}=4$~mT, since that field is found not to affect transmon operation in Ref.~\onlinecite{Bera2021}, and the qubit dephasing time is taken as $\TTs=1~\mu$s, which is typical of optimised transmons~\cite{Houck2008}.
Material parameters for aluminium are density $\rho_0=2700~\mathrm{kg~m}^{-3}$ and Young's modulus $E=68$~GPa. ($\dagger$ To properly distinguish between CSL and purely thermal decoherence, device B requires a smaller $|\alpha|$ which can be achieved by reducing $B_{||}$ by a factor of $\sim$400 to $\sim$0.01~mT.)}

\makebox[\textwidth][c]{
    \begin{ruledtabular}
\begin{tabular}{llll}
\textbf{Parameter} \hspace{6.5cm}				&	&	\textbf{Device A}						&	\textbf{Device B}	\\
\hline
Beam width &$D$ (m)						& $1 \times 10^{-6}$				& $1 \times 10^{-5}$		\\
Beam length &$\ell$ (m)					& $1.9 \times 10^{-3}$			& $8 \times 10^{-2}$		\\
Mass &$m$ (kg)							& $5.1 \times 10^{-12}$			& $2.2 \times 10^{-8}$	\\
Frequency  &$\Omega/2\pi$ (Hz)			& 1400						& 8.1					\\
Zero-point amplitude &$\XZP$ (m)			& $3.4 \times 10^{-14}$			& $6.9 \times 10^{-15}$	\\
Electromechanical coupling  &$|\lambda_\mathrm{on}|/2\pi$ (Hz)	
                                        & $7.2 \times 10^{5}$	& $6.2 \times 10^6$		\\
Grating parameter &$|\alpha|$			& 4.5			& 39 (0.1$^\dagger$)					\\
Maximum separation after single grating  &$\Delta \Xmax$ (m)
										& $6.1 \times 10^{-13}$		& $1.1 \times 10^{-12}$	\\
&&&\\
Energy of gravitational incompatibility (Eq.~\eqref{eq:EG}) &$\EG$ (eV)
                                        & $3.8\times10^{-14}$			& $1.6 \times 10^{-10}$	\\
Corresponding gravitational lifetime &$\tDP$ (s)
                                        & $1.8 \times 10^{-2}$			& $4.2 \times 10^{-6}$	\\
Energy of gravitational incompatibility (Eq.~\eqref{eq:EG_A}) &$\EG$ (eV)
                                        & $3.4\times10^{-15}$			& $7.0 \times 10^{-11}$	\\
Corresponding gravitational lifetime &$\tDP$ (s)
                                        & $1.9 \times 10^{-2}$			& $9.4 \times 10^{-6}$	\\
Energy of gravitational incompatibility (Eq.~\eqref{eq:EG_B}) &$\EG$ (eV)
                                        & $1.4\times10^{-19}$			& $1.9 \times 10^{-15}$	\\
Corresponding gravitational lifetime &$\tDP$ (s)		& $4.7 \times 10^{3}$						& $3.4 \times 10^{-1}$	\\
\label{tab:parameters}
\end{tabular}
    \end{ruledtabular}
}
\end{center}
\end{table*}%

We consider two realisations of the device in Fig.~\ref{fig:device}, with different levels of ambition.
The parameters of these devices are shown in Table~\ref{tab:parameters}.
Device A is designed to have the same mass as the vibrating mirror in the Marshall-Simon-Penrose-Bouwmeester proposal to test for gravitational collapse~\cite{Marshall2003}.
Device B is designed to have the Planck mass.
The dimensions of Devices A and B are chosen to be within the capabilities of microfabrication; obviously, other combinations of length and width could be chosen to give the same mass.
For both devices, the mechanical quality factor is taken to be $Q = 1.1 \times 10^6$, the same as in Ref.~\onlinecite{Bera2021} and typical for aluminium nanobeams.

The qubit frequency is set by the geometry (which determines $\EC$) and the Josephson junction parameters (which determine $\EJ$).
We again take parameters close to those experimentally measured in Ref.~\onlinecite{Bera2021} by assuming $\omegaqo/2\pi=8$~GHz and $\omegaq/2\pi=6$~GHz, which requires $\Delta \Phi \approx 0.620 \, \Phi_0$.

\section{Non-Gaussian mechanical state preparation: the action of the ``grating"}
\label{sec:Effect_of_grating}

We now consider the evolution of the resonator following the conditioning step described in Section~\ref{sec:displacemonprotocol}.
The spatial wavefunction of the conditioned state is proportional to the initial wavefunction multiplied by a sinusoid, meaning that the effective grating performs the transformation~\cite{Khosla2018}:
\begin{equation}
	\ket{\psi}_M \rightarrow \Upsilon_{+}\ket{\psi}_M = \mathcal{R}(\Omega t) \cos \left(|\alpha| \frac{X}{\XZP}\right)\ket{\psi}_M.
	\label{eq:gratingtransformation}
\end{equation}
Here $\ket{\psi}_M$ is the mechanical wavefunction, $\Upsilon_{+}$ is a measurement operator, $\mathcal{R}(\theta) \equiv e^{-i \theta a^\dagger a}$ is the resonator rotation operator, and
\begin{equation}
	\alpha \equiv i \int e^{i\Omega t'} \lambda(t') dt'.
	\label{eq:defineAlpha}
\end{equation}
For a derivation of Eq.~\eqref{eq:gratingtransformation}, see our previous proposal~\cite{Khosla2018}.
Multiplying the wavefunction by a sinusoid is equivalent to dividing it into two branches, one of which has received an impulse $+ \hbar |\alpha| / \XZP$ and the other which has received $-\hbar |\alpha| / \XZP$.
After a quarter of a mechanical period, the centroids of these two branches are spatially separated by a distance
\begin{align}
	\Delta X 	&= \frac{2 \hbar |\alpha|}{m\Omega \XZP}	\\
			&= 4 |\alpha| \XZP~.
\end{align}

The largest separation will occur if $|\alpha|$ is maximised. In Eq.~\eqref{eq:defineAlpha}, the time integral extends at most over the qubit decoherence time $\TTs$, since otherwise the qubit fringes required to create the effective grating are washed out.
Thus, during the qubit precession step, if $\lambda(t)$ is set to its largest accessible value $\lambda_\mathrm{max}$, then the largest accessible separation is
\begin{equation}
	\Delta \Xmax = 4\TTs \lambda_\mathrm{on} \XZP~.
\end{equation}
Taking $\Delta \Phi = 0.620 \Phi_0$ during this step, as explained in Section~\ref{sec:PhysicalParameters}, gives 
\begin{equation}
	\lambda_\mathrm{on} = - 1.16 \times \omegaq \,\frac{\beta_0 \XZP \ell B_{||}}{\Phi_0}~,
\end{equation}
and therefore
\begin{equation}
	\alpha = - 1.16i \times \omegaq \TTs \,\frac{\beta_0 \XZP \ell B_{||}}{\Phi_0}~,
\label{eq:alpha_def}
\end{equation}
and
\begin{align}
	\Delta \Xmax	&= 4.64 \times \omegaq \TTs \,\frac{\beta_0 \XZP^2 \ell B_{||}}{\Phi_0}	\\
						&= 2.32 \times \omegaq \TTs \,\frac{\hbar}{m\Omega} \frac{\beta_0 \ell B_{||}}{\Phi_0}~.
\end{align}

\section{Testing collapse models}

\subsection{Conventional decoherence}
In the standard theory of quantum mechanics, the thermal decoherence of the mechanical state $\rho$ may be modelled by the Caldeira–Leggett master equation
\begin{eqnarray}
\dot{\rho}=-\frac{\mathrm{i}}{\hbar}[H_{0},\rho]-\frac{\mathrm{i\gamma}}{\hbar}[X,\{P,\rho\}]-\mathcal{D}_{\mathrm{th}}[X,[X,\rho]],
\label{eq:thermalME1}
\end{eqnarray}
which describes quantum Brownian motion~\cite{Schlosshauer2007}.
The first term in Eq.~\eqref{eq:thermalME1} describes unitary evolution of the mechanical oscillator with Hamiltonian $H_{0}=\hbar \omegam a^{\dagger}a$,
while the term $-{\mathrm{i\gamma}}[X,\{P,\rho\}]/\hbar$ describes dissipation, which occurs over a timescale $1/\gamma=Q/\omegam$. 
 
The final term $-\mathcal{D}_{\mathrm{th}}[X,[X,\rho]]$ in Eq.~\eqref{eq:thermalME1} describes momentum diffusion and leads to the decoherence of quantum superpositions. Namely, the effect of this term is to cause spatial superpositions of size $2\Delta{X}$ to decohere over a timescale~\cite{zurek1991quantum} 
 \begin{equation}\label{eq:decrate}
    t_{\mathrm{th}}=\dfrac{1}{4\mathcal{D}_{\mathrm{th}}(\Delta X)^2}~.
\end{equation}
Here, the thermal decoherence rate is $\mathcal{D}_{\mathrm{th}}=2 m \gamma k_{\mathrm{B}}T_{\mathrm{eff}}/\hbar^2$ and the thermal occupation of the oscillator is $\bar{N}=k_{\mathrm{B}}T_{\mathrm{eff}}/\hbar\omegam$, which is valid in the limit~\cite{Schlosshauer2007} $k_{\mathrm{B}}T_{\mathrm{eff}}\gg\hbar\omegam$. Here, $T_{\mathrm{eff}}$ is the effective temperature of the environment, which incorporates the cryogenic temperature and other effects such as ambient mechanical noise, which we discuss further in Section \ref{sec:Experimental considerations}.

A term $-\mathcal{D}_{\mathrm{P}}[P,[P,\rho]$ may be added to  Eq.~\eqref{eq:thermalME1} to bring the dynamics into Lindblad form~\cite{breuer2002theory}, where $P$ is the mechanical momentum operator. This term  describes spatial diffusion at a rate $\mathcal{D}_{\mathrm{P}}=\gamma/(8m k_{\mathrm{B}}T_{\mathrm{eff}})$. However, in the temperature regime considered in this work, $k_{\mathrm{B}}T_{\mathrm{eff}}\gg\hbar\omegam$, this term is negligible. We use the corresponding Langevin equations, which describe this quantum Brownian motion in the Heisenberg picture, in order to solve for the open-system dynamics of the mechanical state in Section~\ref{sec:Test_protocol}. 
This enables us to properly model the effects of dissipation and quantum decoherence while avoiding any violation of positivity when solving Eq.~\eqref{eq:thermalME1} at short times with a non-negligible spatial diffusion term~\cite{diosi1993high}.

\subsection{\Diosi-Penrose collapse}
\begin{figure}
	\centering
	\includegraphics[width=\linewidth]{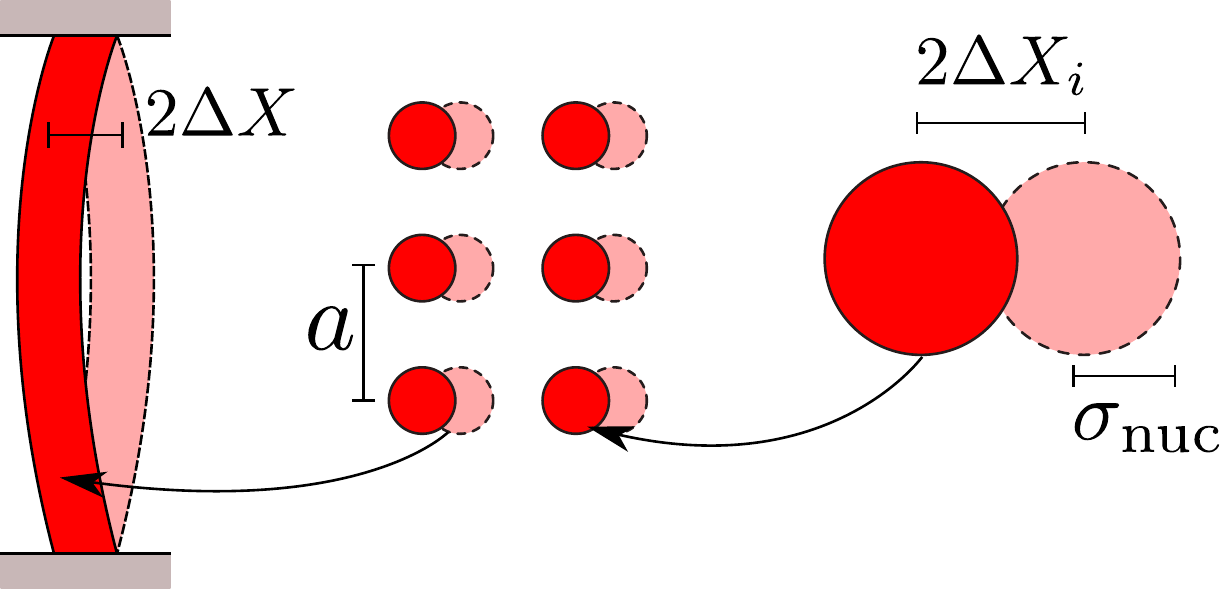}
	\caption{Cartoon interpretation of the displacemon resonator in a spatial superpositon of size $2\Delta X$. Here, $\Delta{X}$ is the root-mean-square displacement of the beam from its equilibrium position. The resonator is a cuboid of volume $V=\ell D^2$ with inter-atomic distance given by $a=2.9 \times 10^{-10}$m for aluminium. Since the majority of an atom's mass resides in the nucleus, the gravitational radius $\sigma_{\mathrm{nuc}}$ labels the nuclear radius (although this parameter is subject to debate; see Section~\ref{sec:sigma}). 
		\label{fig:Penrose_superposition}
	}
\end{figure}

In \Diosi-Penrose theory~\cite{Diosi1989,Penrose1996}, the collapse of a superposition arises from the difference in the mass distributions $\rho(\mathbf{r})$ and $\rho'(\mathbf{r})$ in the two branches. The lifetime of a superposition is
\begin{equation}\label{eq:dprate}
	 \tDP = \frac{\hbar}{\EG},
\end{equation}
where~\cite{Howl2019}
\begin{equation}
	\EG \equiv 4\pi G \gammaDP \iint \frac{[\rho(\mathbf{r}) - \rho'(\mathbf{r})] [\rho(\mathbf{r'}) - \rho'(\mathbf{r'})]}{|\mathbf{r}-\mathbf{r'}|}\, d^3 \mathbf{r} \, d^3 \mathbf{r'}
	\label{eq:EGint}
\end{equation}
parameterizes the incompatibility of the space-time generated by the two branches of the superposition, and $\mathbf{r}$ and $\mathbf{r'}$ are spatial coordinates.
Equation~\eqref{eq:EGint} is proportional to the gravitational self-energy of the difference between the two mass distributions, and $\gammaDP$ is a constant~\cite{Howl2019} which takes the value $\gammaDP=1/8\pi$ if $|\EG|$ is equal to the gravitational self-energy (rather than only proportional to it).

For the devices considered here, the spatial separation $2\Delta X$ is much smaller than the distance between adjacent atoms $a$.
(Compare Table~\ref{tab:parameters} with the nearest-neighbour distance in aluminium, which is $a=2.9 \times 10^{-10}$~m.)
This means that Eq.~\eqref{eq:EGint} can be broken down into a sum of self-energies created by the individual nuclei, each in a superposition:
\begin{equation}
	\EG \approx \sum_i \EG^{(1)}(2\Delta X_i)~.
	\label{eq:EG2}
\end{equation}
Here $\EG^{(1)}(2\Delta X_i)$ is the self-energy of the superposition of a single nucleus separated by $2\Delta X_i$, which is the separation of the entire beam at the location of the $i^{\mathrm{th}}$ nucleus (see Fig.~\ref{fig:Penrose_superposition}).
Since
\begin{equation}
	X_i (Z,t) = X(t) u_0(Z_i),
\end{equation}
with $u_0(Z)$ defined by Eq.~\eqref{eq:u0}, Eq.~\eqref{eq:EG2} becomes
\begin{equation}
	\EG(t) = 8\pi\gammaDP \frac{D^2 \rho_0}{\mA} \int \EG^{(1)} \left(2 X(t) u_0(Z)\right) \, dZ
\end{equation}
where $\rho_0$ and $\mA$ are the density and atomic mass of the material from which the beam is made.
The factor~2 takes account of the fact that the two branches of the wavefunction correspond to beam displacements in different directions.

A full expression for $\EG^{(1)}$ is given in Ref.~\onlinecite{kleckner2008creating}; modelling the nuclei as hard spheres with radius $\sigma$, it is
\begin{equation}
	\EG^{(1)}(2\Delta X_i) = 8\pi\gammaDP \, G\mA^2 \left( \frac{6}{5\sigma}-\frac{1}{2\Delta X_i}\right).
	\label{eq:EGhowl}
\end{equation}
The value of $\sigma$ is the size of the nuclear mass distribution that acts as the gravitational source.
The simplest choice is that $\sigma$ is the nuclear radius, which in aluminium~\cite{Angeli2013} is $\sigma_\mathrm{nuc}=3.06 \times 10^{-15}$~m.
Since Table~\ref{tab:parameters} shows that $\Delta X \gg \sigma_\mathrm{nuc}$, the first term in Eq.~\eqref{eq:EGhowl} dominates, meaning that
\begin{equation}
	\EG^{(1)} \approx 8\pi\gammaDP \,\frac{6}{5}\frac{ G \mA^2}{\sigma_\mathrm{nuc}}
	\label{eq:EG1}
\end{equation}
and therefore
\begin{align}
	\EG	&= 8\pi\gammaDP \, \frac{6}{5} \frac{G D^2\ell \rho_0 \mA}	{\sigma}    \\
	    &=  8\pi\gammaDP \, \frac{6}{5} \frac{G m \mA}{\sigma_\mathrm{nuc}}     \\
		&=  \frac{6}{5} \frac{G m \mA}{\sigma_\mathrm{nuc}}		\label{eq:EG}
\end{align}
where $m$ is the mass of the beam, and the last line makes the assumption~\cite{Howl2019} that $\gammaDP=1/8\pi$.
This leads to the predicted collapse lifetime $\tDP$ shown in Table~\ref{tab:parameters}.
Any superposition in device A would collapse due to gravity in approximately 18~ms, or 25 mechanical periods; for device B, the collapse time is much less than the mechanical period, so that a superposition that survives even for half a period would be enough to exclude this collapse model.

For gravitational collapse to be observable, the conventional decoherence rate needs to be reduced below the gravitational rate $1/\tDP$. In Section~\ref{sec:Test_protocol}, we outline a scheme to detect this decoherence rate using the displacemon device.

\subsubsection{Alternative choices for the gravitational radius $\sigma$}
\label{sec:sigma}

Above, we assumed that $\sigma$ is the same as the nuclear radius $\sigma_\mathrm{nuc}$.
It has also been suggested~\cite{kleckner2008creating} that $\sigma$ should be the radius of the ground-state wavepacket if the mechanical state is initialized in the ground state prior to superposition state preparation.
There are two ways of defining this.
One is to say that the wavepacket size is equal to the beam's zero-point amplitude, i.e. $\sigma = \XZP$.
If we make the crude approximation that the wavepacket gravitates in the same way as a uniform sphere with this radius, then we still have $\Delta \Xmax \gg \sigma$ and so Eq.~\eqref{eq:EG} is modified to
\begin{equation}
    \EG = \frac{6}{5} \frac{Gm \mA}{\XZP}.
    \label{eq:EG_A}
\end{equation}
Alternatively, the wavepacket size may be taken as the zero-point amplitude of a single atom in the crystal, which is
\begin{align}
	\sigma = \XZP^{(1)}	&= \sqrt{\frac{\hbar}{2 \mA \Omega_\mathrm{p}}}	\\
			&\approx 8 \times 10^{-12}~\mathrm{m} ~\text{in aluminium}
\end{align}
where $\Omega_\mathrm{p}$, the maximum of the lower phonon dispersion branch, which in aluminium~\cite{Stedman1966}, is $\Omega_\mathrm{p} \approx 2\pi \times 3$~THz.
We now have $\Delta \Xmax < \sigma$, and so (in the approximation that the wavepacket still gravitates as a uniform sphere), Eq.~\eqref{eq:EGhowl} is modified to
\begin{equation}
    \EG^{(1)}\left(2\Delta X_i\right) \approx 8\pi\gammaDP \, \frac{ {{2}} G \mA^2 (\Delta X_i)^2}{\left(\XZP^{(1)}\right)^3}
\end{equation}
which follows an expression from Ref~\onlinecite{Howl2019}, keeping only the leading term in $\Delta X_i/\XZP^{(1)}$.
A full analysis should take into account that $\Delta X_i$ depends on both $Z$ and $t$, but in the gross approximation that $\Delta X_i \sim \Delta \Xmax$ then we find
\begin{equation}
    \EG \sim {2} G m \mA \frac{(\Delta \Xmax)^2}{\left(\XZP^{(1)}\right)^3}
    \label{eq:EG_B}
\end{equation}
again assuming $\gammaDP = 1/8\pi$.

For completeness the gravitational energies and lifetimes that follow from Eqs.~\eqref{eq:EG_A} and~\eqref{eq:EG_B} are listed in Table~\ref{tab:parameters}.

\subsubsection{Comparison to thermal decoherence rate}

To observe the collapse predicted by the DP model we require that $t_{\mathrm{th}}>\tDP$. Using Eqs.~\eqref{eq:decrate} and \eqref{eq:dprate} this condition may be rewritten as 
\begin{eqnarray}\label{eq:decvsdp}
\bar{N}<\frac{\EG Q}{{{8m\omegam^2}}(\Delta X)^2}~.
\end{eqnarray}
The right-hand side of Eq.~\eqref{eq:decvsdp} is a function of the separation $\Delta X$. 
Note that $\Delta X$ is a dynamical variable that varies between an initial value of zero and $\Delta{X}_{\mathrm{max}}$ over one quarter of a mechanical cycle. Over this timescale, the dissipation rate is much smaller than the mechanical frequency $\omegam/2\pi$ and the decoherence rates, so may be ignored. Furthermore, the separation $\Delta{X}_{\mathrm{max}}$ may be controlled with the magnetic field $B_{||}$. Therefore, we may maximize  Eq.~\eqref{eq:decvsdp} with respect to $\Delta X$ to estimate the maximum thermal occupation of the environment $\bar{N}^{\mathrm{max}}$ that still allows DP collapse to be observed, and the corresponding environmental temperature $T_{\mathrm{eff}}^{\mathrm{max}}$.

To maximize the right-hand side of Eq.~\eqref{eq:decvsdp} we account for the dependence of $\EG$ on $\Delta{X}$ through  Eq.~\eqref{eq:EGhowl} in the approximation $\Delta{X}_{i}\approx\Delta{X}$, which is valid for $\ell\gg D$. The term $\EG/(\Delta X)^2$ is then a decreasing function of $\Delta X$ for~\cite{Howl2019} $\Delta{X}\geq\sigma$. Hence, in the region $\Delta{X}\geq\sigma$ we find that Eq.~\eqref{eq:decvsdp} is maximized at $\Delta{X}=\sigma$ to give $\bar{N}<\bar{N}^{\mathrm{max}}$ with
\begin{eqnarray}
\bar{N}^{\mathrm{max}}=7G\mA Q/(80\sigma^3\omegam^2)~.
\end{eqnarray}
Firstly, we take the gravitational radius to be the nuclear radius $\sigma=\sigma_{\mathrm{nuc}}$, in which case Device A requires $\bar{N}^{\mathrm{max}}=1.3\times10^5$ ($T_{\mathrm{eff}}^{\mathrm{max}}=8.7\,\mathrm{mK}$) while Device B requires $\bar{N}^{\mathrm{max}}=3.9\times10^9$ ($T_{\mathrm{eff}}^{\mathrm{max}}=1.5\,\mathrm{K}$). Secondly, we take $\sigma$ to be the zero-point amplitude of the beam $\sigma=\XZP$. In this case we find a more stringent requirement for Device A of $\bar{N}^{\mathrm{max}}=95$ ($T_{\mathrm{eff}}^{\mathrm{max}}=6.3\,\mu\mathrm{K}$), while Device B may operate at $\bar{N}^{\mathrm{max}}=3.4\times10^8$ ($T_{\mathrm{eff}}^{\mathrm{max}}=0.13\,\mathrm{K}$). Device B can operate at a higher effective temperature for both choices of $\sigma$, which is attributed to the $1/\omegam^2$ dependence of $\bar{N}^{\mathrm{max}}$. These relatively high effective temperatures required by Device B are encouraging given recent experimental progress in cooling Planck mass objects~\cite{neuhaus2021laser}.

Finally, we consider the case when the gravitational radius is given by the zero-point amplitude of a single atom in the crystal $\sigma=\XZP^{(1)}$. As this corresponds to the region $\Delta{X}<\sigma$ we take $\Delta{X}=\Delta{X}_{\mathrm{max}}$ to provide an estimate for $\bar{N}^{\mathrm{max}}$. 
Combining Eqs.~\eqref{eq:EG_B} and \eqref{eq:decvsdp} then gives $\bar{N}<\bar{N}^{\mathrm{max}}=G\mA Q/\left({4}\omegam^2(\XZP^{(1)})^3\right)$, which gives $\bar{N}^{\mathrm{max}}={2.1}\times10^{-5}$ ($T_{\mathrm{eff}}^{\mathrm{max}}={1.4}\times10^{-12}\,\mathrm{K}$) for Device A and $\bar{N}^{\mathrm{max}}={0.62}$ ($T_{\mathrm{eff}}^{\mathrm{max}}={2.4}\times10^{-10}\,\mathrm{K}$) for Device B. These results break our initial assumption that $\kB \Teff\gg\hbar\omegam$ and do not correspond to physically realizable temperatures. Hence the displacemon is not capable of sensing DP collapse with $\sigma=\XZP^{(1)}$.

Treating the constituents as uniform spheres~\cite{Howl2019}, the term $\EG/(\Delta{X})^2$ is maximized in the limit $\Delta{X}\rightarrow0$, which seems to increase the $\bar{N}$ our devices can tolerate in \eqref{eq:decvsdp}.  
However, we note that in this limit, both decoherence timescales $\tDP$ and $t_{\mathrm{th}}$ tend to infinity and we do not expect to see any collapse as there is no spatial superposition. Furthermore, in our testing protocol---see Section~\ref{sec:Test_protocol}---in the limit $\Delta{X}\rightarrow0$ the mechanical state is initialized in a thermal state. In this limit, there is no quantum decoherence and the thermal state simply equilibrates with its environment.

\subsection{Continuous Spontaneous Localisation (CSL)}
In order to model the dynamical evolution of the displacemon resonator under the assumptions of CSL, we can add the term $-\mathcal{D}_{\mathrm{CSL}}[X,[X,\rho]]$ to the right-hand side of the master equation in Eq.~\eqref{eq:thermalME1}, corresponding to a momentum diffusion process with rate $\mathcal{D}_{\mathrm{CSL}}$. This additional term is valid for regimes where the resonator has an oscillation amplitude~\cite{Nimmrichter2014} $X_{\mathrm{ZP}}\ll r_{\mathrm{CSL}}$, and would only be detectable over conventional thermal noise if $\mathcal{D}_{\mathrm{CSL}}>\mathcal{D}_{\mathrm{th}}$. The geometry of the superposed macroscopic object dictates $\mathcal{D}_{\mathrm{CSL}}$. Assuming that the superposed displacemon beam is a cuboid of length $\ell$ and width $D$ such that the volume is given by $V=\ell D^2$, it follows that~\cite{Nimmrichter2014,Ferialdi2020}:
\begin{equation}
\begin{split}
    \mathcal{D}_{\mathrm{CSL}}&=\lambda_{\mathrm{CSL}} \frac{N_{\mathrm{n}}^2}{D^2}\Gamma_1\bigg(\frac{D}{\sqrt{2}r_{\mathrm{CSL}}}\bigg)\Gamma_1\bigg(\frac{\ell}{\sqrt{2}r_{\mathrm{CSL}}}\bigg)\\
    &~~~~~~~~\times[1-e^{-D^2/4r^2_{\mathrm{CSL}}}]~,
\end{split}
\end{equation}
where
\begin{equation}
    \Gamma_1(x)=\frac{2}{x^2}\bigg[ e^{-x^2/2}-1+\sqrt{\frac{\pi}{2}} x \mathrm{erf}\bigg(\frac{x}{\sqrt{2}}\bigg)\bigg]~,
\end{equation}
and $N_{\mathrm{n}}$ (not to be confused with $\bar{N}$) is the total number of nucleons that make up the resonator. Using $N_{\mathrm{n}}=m/\mathrm{amu}$, we have that $N_{\mathrm{n}}=3.1\times10^{15}$ for Device A, and $1.3\times10^{19}$ for Device B.
The evolution of the resonator is governed by Eq.~\eqref{eq:thermalME1} (adapted to include CSL decoherence, as discussed) and is captured by the following quantum Langevin equations:
\begin{subequations}
    \begin{align}
        \dot{X}&=P/m\\
      \dot{P}&=-m\omegam^2 X-\gamma P + f_{\mathrm{th}}(t)+f_{\mathrm{CSL}}(t)~,
\end{align}
\end{subequations}
where $f_{\mathrm{th}}$ and $f_{\mathrm{CSL}}$ are the stochastic force terms arising from the standard theory of thermal decoherence, and CSL, respectively. These operators obey the following relations:
\begin{subequations}
\begin{align}
        \braket{f_{\mathrm{th}}(t)f_{\mathrm{th}}(t')}&=\hbar^2 \mathcal{D}_{\mathrm{th}}\delta (t-t')\\
        \braket{f_{\mathrm{CSL}}(t)f_{\mathrm{CSL}}(t')}&=\hbar^2 \mathcal{D}_{\mathrm{CSL}}\delta(t-t')~
    \end{align}
\end{subequations}
and $\braket{f_{\mathrm{th}}(t)f_{\mathrm{CSL}}(t')}=0$ at all times since the two noise sources are uncorrelated. By exploiting qubit measurements on the displacemon device, we can test the thermal decoherence of a mechanical resonator and deduce whether the rate agrees with conventional decoherence theory or whether there is a deviation which can be attributed to a non-zero $\mathcal{D}_{\mathrm{CSL}}$. The proposed procedure for this is described in detail in the next section.

\section{Testing protocol}
\label{sec:Test_protocol}
\floatsetup[figure]{style=plain,subcapbesideposition=top}
\begin{figure*}
\centering
  \sidesubfloat[]{\includegraphics[
  height=5.2cm]{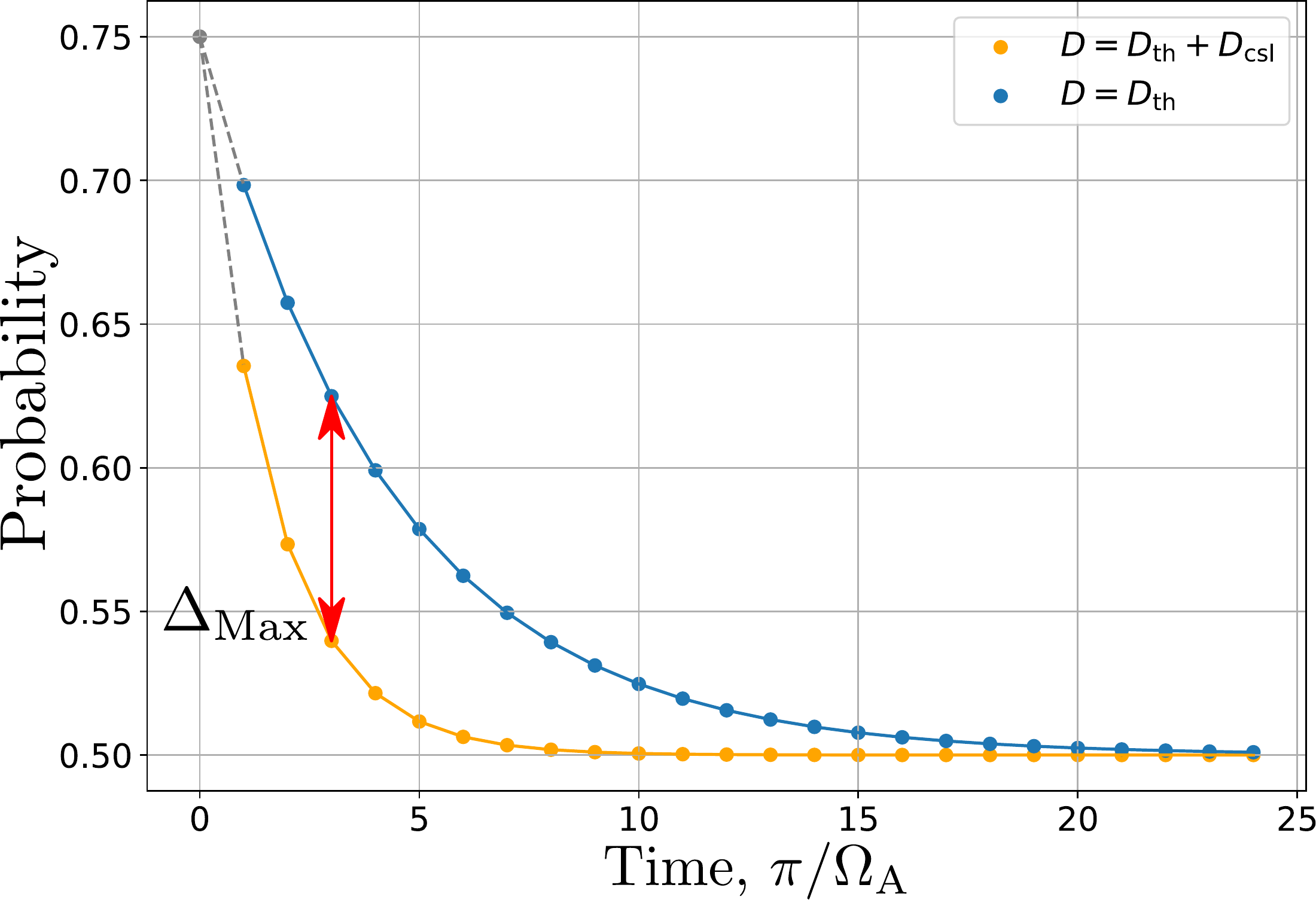}
  \label{fig:decay_plots_A}}
$\qquad$
  \sidesubfloat[]{\includegraphics[height=5.2cm]{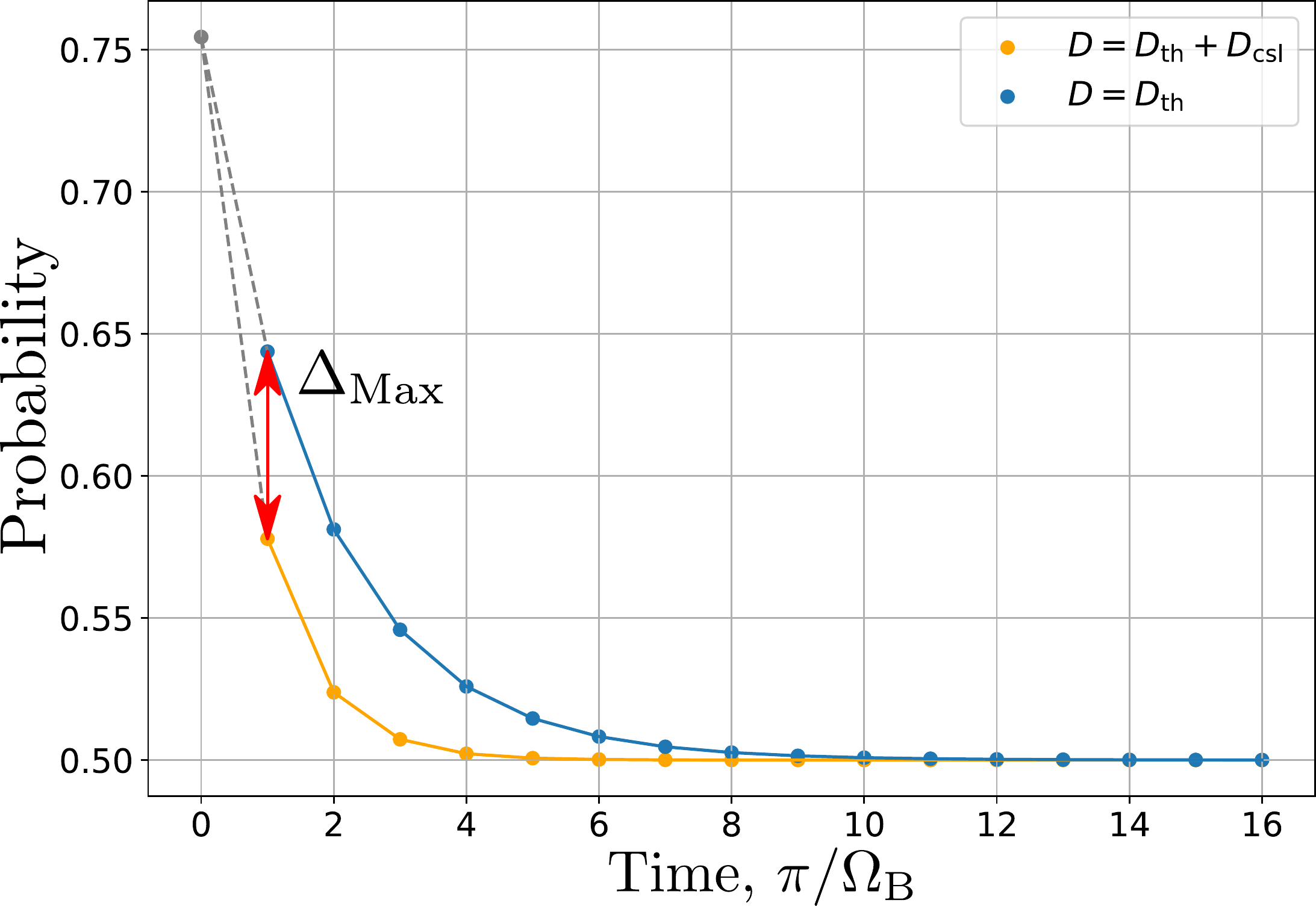}
  \label{fig:decay_plots_B}}%
  \caption{Probability of measuring the displacemon qubit in the $\ket{+}$ state, given it was initially measured in $\ket{+}$, as a function of the the number of mechanical half-cycles $k$ elapsed between the two effective gratings (see Eq.~\eqref{eq:probability_decay}). The blue line shows the probability decaying as predicted by standard quantum mechanics with momentum diffusion rate given by $\mathcal{D}=\mathcal{D}_{\mathrm{th}}$. The orange line has incorporated the CSL decoherence rate giving a total momentum diffusion rate of $\mathcal{D}=\mathcal{D}_{\mathrm{th}}+\mathcal{D}_{\mathrm{CSL}}$. The red arrow indicates the position of maximum percentage difference $\Delta_{\mathrm{Max}}$ we can expect between the two theories for this parameter set; these points correspond to the orange dots in Figs.~\ref{fig:contour_plots_A} - \ref{fig:contour_plots_B}. (a) Plot for Device A using the values in Table~\ref{tab:parameters}, $\lambda_{\mathrm{CSL}}=10^{-11}$~Hz and $r_{\mathrm{CSL}}=10^{-7}$~m, initial mechanical occupation $\bar{n}=100$, and a bath occupation number $\bar{N}=10^{3}$ (or~0.07~mK). (b) Plot for Device B, again using values from Table~\ref{tab:parameters}, $\lambda_{\mathrm{CSL}}=10^{-11}$~Hz,  $r_{\mathrm{CSL}}=10^{-7}$~m, $\bar{n}=100$, and $\bar{N}=5\times 10^6$ (or 1.9~mK).}
  \label{fig:decay_plots}
\end{figure*}

\floatsetup[figure]{style=plain,subcapbesideposition=top}
\begin{figure*}
\centering
  \sidesubfloat[]{\includegraphics[width=0.49\linewidth ]{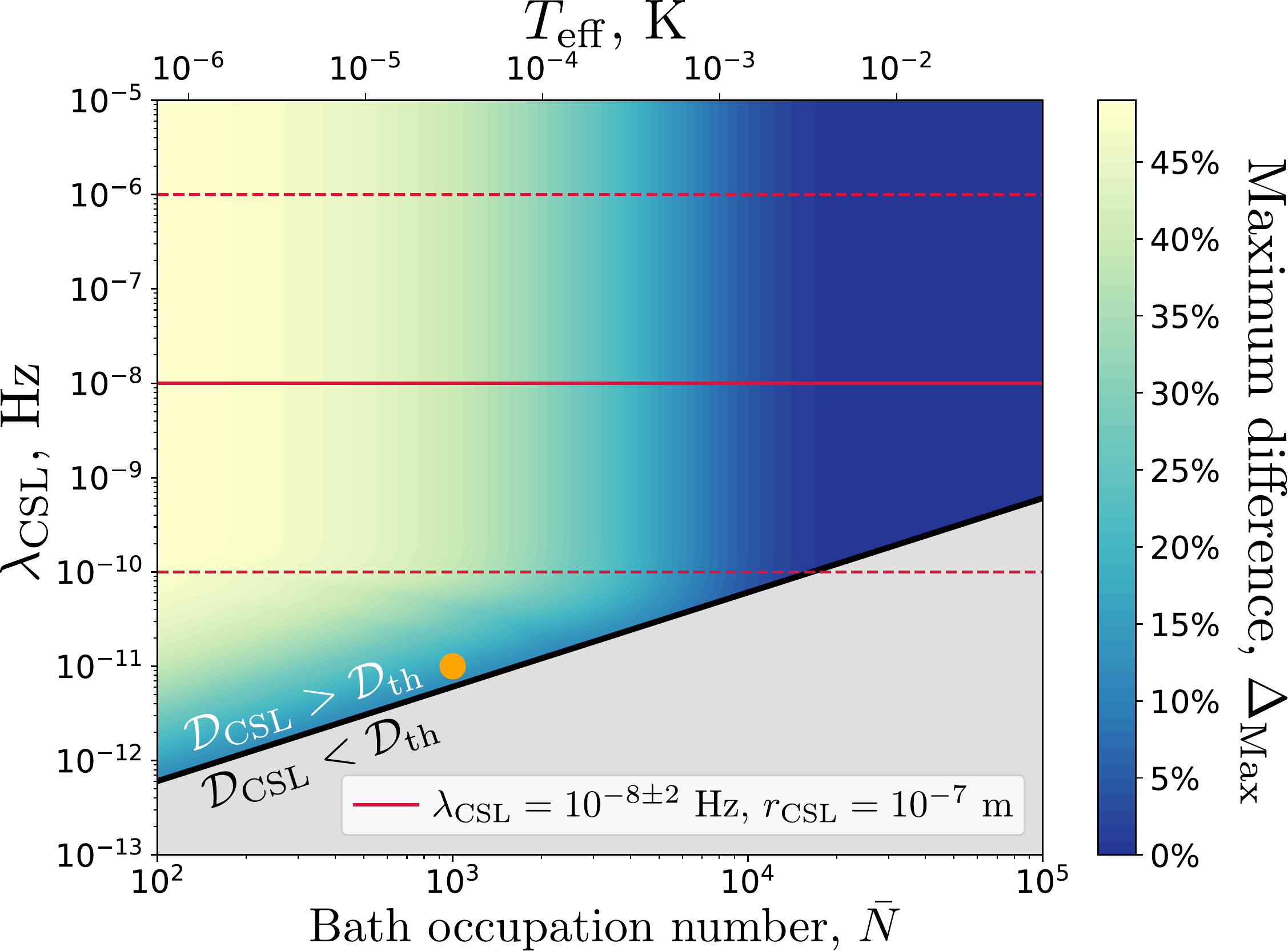}
  \label{fig:contour_plots_A}}
  \sidesubfloat[]{\includegraphics[
  width=0.49\linewidth]{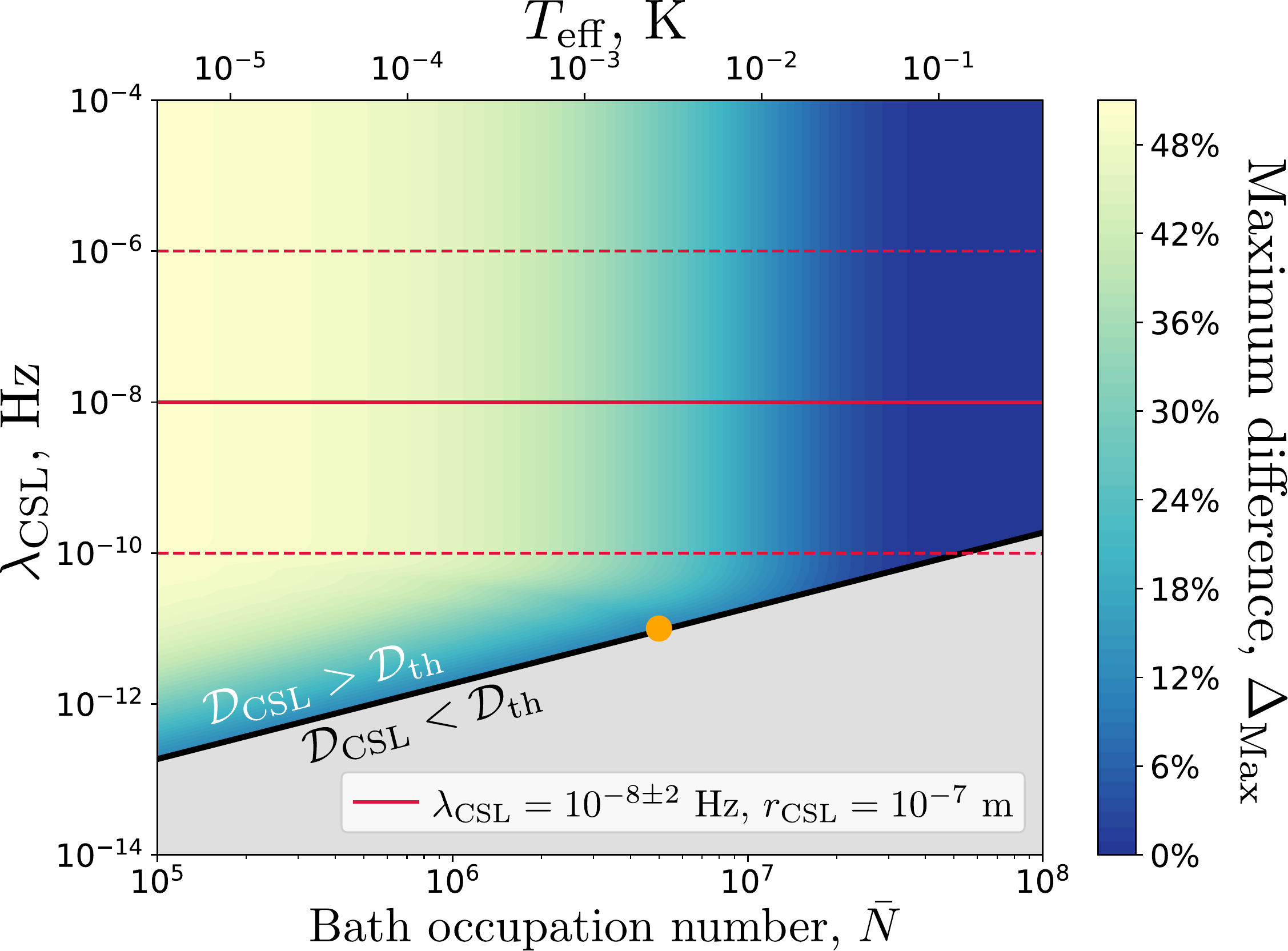}
  \label{fig:contour_plots_B}}%
  \caption{Maximum percentage difference $\Delta_{\mathrm{Max}}$ between the probabilities predicted by standard quantum mechanics versus CSL theory, found by maximising Eq.~\eqref{eq:prob_diff} with respect to $k$. Results are plotted as a function of $\lambda_{\mathrm{CSL}}$ and bath thermal occupation number $\bar{N}$, with the equivalent effective temperature $T_{\mathrm{eff}}$  shown along the top axis. (a) Plot for Device A using parameters in the corresponding column of Table \ref{tab:parameters}. (b) Plot for Device B, again using parameters in the corresponding column of Table \ref{tab:parameters}.
  For both plots we use $r_\mathrm{CSL}=10^{-7}$~m and assume the mechanical resonator has been pre-cooled to a thermal state with initial occupation number $\bar{n}=100$. The light yellow indicates regions of parameter space which we expect to be most sensitive to CSL effects as they exhibit the greatest deviation from standard quantum theory. The grey exclusion region, demarcated by the black line, indicates that $\mathcal{D}_{\mathrm{CSL}}<\mathcal{D}_{\mathrm{th}}$, meaning that the effects of CSL cannot be detected above thermal decoherence. The red horizontal lines show Adler's suggested theoretical parameters for $\lambda_{\mathrm{CSL}}$ (and $r_{\mathrm{CSL}}$). The orange dots correspond to the parameters used in Figs. \ref{fig:decay_plots_A}-\ref{fig:decay_plots_B} which explicitly show the decay of the probabilities $P(k,\mathcal{D}_{\mathrm{th}})$ and $P(k,\mathcal{D}_{\mathrm{th}}+\mathcal{D}_{\mathrm{CSL}})$ with time.}
  \label{fig:contour_plots}
\end{figure*}

The displacemon device has several features which can be exploited to dynamically quantify the degree of decoherence of the resonator. From the statistics of qubit measurements taken at regular time intervals one can infer both the extent of decoherence in the mechanical system and the rate at which this occurs. The following method should be read with CSL in mind as it exploits the dynamical description provided by the Langevin equations (Eqs. (\ref{eq:Langevin_solution_X} - \ref{eq:Langevin_solution_P})). While Penrose does not offer a dynamical description for gravity-induced collapse, we shall demonstrate that the following protocol can still be used to determine whether or not total decoherence of the mechanical state occurs on the timescale of $t_{\mathrm{DP}}$ or $t_{\mathrm{th}}$.

Our method to test the decoherence rate of a mechanical superposition using the displacemon device is as follows:
\begin{enumerate}
    \item The resonator is cooled towards its ground state using the qubit-measurement pre-cooling scheme outlined in Ref. \onlinecite{Khosla2018}. By using a sequence of pulses and selecting specific qubit outcomes, we can discard higher energy states and select the lowest-energy thermal states. Thus, the mechanical state $\rho_{\bar{n}}$ is prepared in a thermal state with $\bar{n}$ being the initial thermal occupation number.
    \item The qubit-mechanical interaction is switched on (Eq.~\eqref{eq:Hamiltonian}) and a measurement of the qubit in the state $\ket{+}$ heralds an effective diffraction grating (see Eq.~\eqref{eq:gratingtransformation}) in the mechanical system. This then leads to a spatial mechanical superposition (as outlined in Section \ref{sec:Effect_of_grating}). The post-measurement state of the mechanical system is given by:
    \begin{equation}
        \rho_{\mathrm{in}}=\frac{\Upsilon_{+} \rho_{\bar{n}} \Upsilon_{+}^\dag}{\Tr[\Upsilon_{+}^\dag \Upsilon_{+} \rho_{\bar{n}}]}~,
    \end{equation}
    where the denominator is the heralding probability of creating such a state: $\Tr[\Upsilon^\dag_{+}\Upsilon_{+}\rho_{\bar{n}}]=(1+e^{-2(1+2\bar{n}|\alpha|^2)})/2$. 
    We neglect any mechanical decoherence which may occur during the implementation of the measurement operator since $\Upsilon_{+}$ is enacted on the timescale of the qubit lifetime $T_2^{*}$ which is much shorter than the mechanical period.
    \item Following the qubit measurement outcome $\ket{+}$, the mechanical state $\rho(t)$ is decoupled from the qubit and the position and momentum operators evolve according to Eqs. \eqref{eq:Langevin_solution_X}-\eqref{eq:Langevin_solution_P}, in the Heisenberg picture. 
    \item Finally, after a time $t_k$, we apply a second effective grating operation $\Upsilon_{+}$ by again switching on the qubit-mechanical interaction and measuring the qubit in the state $\ket{+}$. The time elapsed between applying the first and second gratings is $t_k=k\pi/\omegam$ where $k$ is the integer number of mechanical half-periods. By waiting for mechanical half-periods we can neglect the $R$ operator in Eq.~\eqref{eq:gratingtransformation}. As the qubit is only coupled to the mechanical system during the fast grating operations, we are not constrained by the qubit coherence time; a key strength of our protocol. Over many experimental runs, we calculate the probability of measuring the qubit in the $\ket{+}$ state at time intervals $t_k$ after it was first measured in the $\ket{+}$ state. This conditional probability is denoted by $P(k,\mathcal{D})=\Tr[\Upsilon^\dag_{+}\Upsilon_{+}\rho(t_k)]$, where $\mathcal{D}=\mathcal{D}_{\mathrm{th}}$ according to conventional decoherence theory, or $\mathcal{D}=\mathcal{D}_{\mathrm{th}}+\mathcal{D}_{\mathrm{CSL}}$ if we incorporate the predictions from CSL theory. In the absence of any decoherence effects ($\mathcal{D}=0$) this expression can be readily evaluated to give:
    \begin{equation}
        P(k,0)=\frac{3+4e^{-2(1+2\bar{n})|\alpha|^2}+e^{-8(1+2\bar{n})|\alpha|^2}}{4(1+e^{-2(1+2\bar{n})|\alpha|^2})}~,
    \end{equation}
    and is bounded between [0.75, 1] for all $\bar{n}$ and $|\alpha|$. Therefore, observing $P(k,\mathcal{D})<0.75$ necessarily implies the presence of decoherence either by conventional means, or via an additional collapse-theory mechanism. If $P(k,\mathcal{D})=0.5$ then the qubit is equally likely to be in the $\ket{+}$ or $\ket{-}$ state and so the mechanical state has completely decohered. 
    \item Alternatively, to test DP-style collapse we can measure the timescale required for $P(k,\mathcal{D})$ to reduce to an appropriate value (depending on the initial state) indicating that the mechanical state has undergone decoherence, and compare this with the timescales $t_{\mathrm{DP}}$ and $t_{\mathrm{th}}$. 
    \item Note: We are interested in the probability that a second qubit measurement yields $\ket{+}$. As this is a binary-outcome measurement, instances where the second measurement is $\ket{-}$ also contribute to our empirical estimate of $P(k,\mathcal{D})$ and so these experimental runs need not be discarded.
\end{enumerate}
\floatsetup[figure]{style=plain,subcapbesideposition=top}

Using this experimental recipe we can investigate how the probability $P(k,\mathcal{D})$ scales with various parameters as we tune $\mathcal{D}_{\mathrm{CSL}}$ and $\mathcal{D}_{\mathrm{th}}$. From this we can identify the regions of parameter space where the predicted effects of CSL are greater than those of conventional decoherence and thus can be feasibly tested. 

We can derive an analytical expression for $P(k,\mathcal{D})$ by recasting it in the Heisenberg picture $P(k,\mathcal{D})=\braket{\Upsilon^\dag_{+}(t_k)\Upsilon_{+}(t_k)}$, where $\Upsilon_{+}(t_k)=\cos (|\alpha| X(t_k)/X_{\mathrm{ZP}})$. By inserting the solutions to the Langevin equations into the definition of $\Upsilon_{+}(t_k)$ (see Eq.~\eqref{eq:gratingtransformation}) we expect the probability to decay with $k$ according to:
\begin{equation}
P(k,\mathcal{D})=\frac{1}{2}+\frac{1}{2}P_0(t_k)\exp\bigg[-8\mathcal{D}|\alpha|^2\frac{(1-e^{-k\pi\gamma/\omegam})}{\gamma(4+\gamma/\omegam)}\bigg]~.
\label{eq:prob_full}
\end{equation}
For a derivation of this formula see Appendix \ref{app:CSL}. In regimes, such as ours, with high quality factors $Q=\omegam/\gamma \gg 1$ we can make the following approximation
\begin{equation}
    P(k,\mathcal{D})\approx\frac{1}{2}+\frac{1}{2}P_{0}(k)\exp \bigg( -2\mathcal{D}|\alpha|^2 \frac{k\pi}{\omegam}\bigg)~,
    \label{eq:probability_decay}
\end{equation}
demonstrating that the probability decays exponentially with time towards a value of $0.5$. The slowly-decaying envelope function $P_0(k)$ (defined in Eq.~\eqref{eq:P0} of Appendix~\ref{app:CSL}) determines the initial probability $P(0,\mathcal{D})$ and depends on $k$, the grating parameter $|\alpha|$, and the properties of the mechanical oscillator $\gamma$, $\omegam$, and $\bar{n}$. 

The theoretical decay of the probabilities $P(k,\mathcal{D}_{\mathrm{th}})$ and $P(k,\mathcal{D_{\mathrm{th}}}+\mathcal{D}_{\mathrm{CSL}})$ are plotted for Devices A and B in Figs.~\ref{fig:decay_plots_A} and \ref{fig:decay_plots_B}. For these plots, the CSL decay rate has been calculated assuming $r_{\mathrm{CSL}}=10^{-7}$m and $\lambda_{\mathrm{CSL}}=10^{-11}$~Hz (a factor of $10^{3}$ times smaller than Adler's predicted value~\cite{Adler2007}, thus probing a more challenging region of parameter space). The mechanical parameters and $|\alpha|$ have been set to the values given in Table~\ref{tab:parameters}. The probability in Eq.~\eqref{eq:probability_decay} decays exponentially with $|\alpha|^2$ and measurements are only performed after at least half a mechanical period has elapsed. Accordingly, we have used $|\alpha|=0.1$ for Device B so that CSL is properly distinguishable from purely thermal decoherence; this can be achieved by setting $B_{||}=0.01$~mT (see Eq.~\eqref{eq:alpha_def}). The initial thermal occupation numbers of the resonators are set to $\bar{n}=100$, assuming both devices have been pre-cooled. The bath environment is assumed to have a thermal occupation number $\bar{N}=k_{\mathrm{B}}T_{\mathrm{eff}}/(\hbar \omegam)$ with values $\bar{N}=10^3$ (equivalent to an effective temperature of 0.07~mK) for Device A and $\bar{N}=5\times 10^6$ (or 1.9~mK) for Device B. 
We have chosen these values of $\bar{N}$ for these plots in order to most clearly visually demonstrate the exponential decay of $P(k,\mathcal{D})$. However, we discuss below the experimental challenge of achieving such low temperatures, and the interplay between the bath temperature and testable CSL parameters. 

To maximally distinguish between decoherence effects in standard quantum theory versus CSL, we can investigate the percentage difference between the expected probabilities of a second $\ket{+}$ qubit measurement as predicted by the two theories:
\begin{equation}
    \Delta=2\frac{P(k,\mathcal{D}_{\mathrm{th}})-P(k,\mathcal{D}_{\mathrm{th}}+\mathcal{D}_{\mathrm{CSL}})}{P(k,\mathcal{D}_{\mathrm{th}})+P(k,\mathcal{D}_{\mathrm{th}}+\mathcal{D}_{\mathrm{CSL}})}~.
    \label{eq:prob_diff}
\end{equation}
For this we use the full analytic result for $P(k,\mathcal{D})$ given in Eq.~\eqref{eq:prob_full}.
As we scan the discrete variable $k$ (i.e. vary the time between applying the two diffraction gratings) we can find the maximum percentage difference between the two probabilities $\Delta_{\mathrm{Max}}$ for different values of $\mathcal{D}_{\mathrm{th}}$ and $\mathcal{D}_{\mathrm{CSL}}$. This quantity is plotted as a function of $\lambda_{\mathrm{CSL}}$ and $\bar{N}$ for Devices A and B in Figs. \ref{fig:contour_plots_A} and \ref{fig:contour_plots_B}, using the parameters in Table \ref{tab:parameters}. As before, we have used initial resonator occupation numbers of $\bar{n}=100$. We only consider $\bar{N}>\bar{n}=100$ as we expect the bath to always be warmer than the device itself. 

For both devices, Fig.~5 shows that one should expect appreciable differences between $P(k, \mathcal{D})$ for the two theories, provided sufficiently low environmental temperatures can be achieved. Device B is more suitable for probing lower $\lambda_{\mathrm{CSL}}$ values at more feasible bath temperatures. One can see that even at $\bar{N}=2\times 10^7$ (a possibly feasible effective temperature of $7.7\,\mathrm{mK}$), we expect Device~B to exhibit at least an $\approx 5\%$ difference between CSL and standard quantum theory for $\lambda_{\mathrm{CSL}}>10^{-10}\,\mathrm{Hz}$. In contrast, much lower bath temperatures are required for Device A to probe similar CSL rates. To confidently distinguish between CSL and standard thermal decoherence we require $\mathcal{D}_{\mathrm{CSL}}>\mathcal{D}_{\mathrm{th}}\propto2\bar{N}\omegam/Q$ and so have only included the region in which this is true. Since $\mathcal{D}_{\mathrm{CSL}}$ is directly proportional to $\lambda_{\mathrm{CSL}}$, the smallest value of $\lambda_{\mathrm{CSL}}$ that we can conclusively probe is proportional to the mechanical frequency $\omegam$, which is much higher for Device A.

While we have focused on the displacemon device, we note that in principle our testing protocol is applicable to other optomechanical platforms that achieve an equivalent measurement operator to Eq.~\eqref{eq:gratingtransformation}. For example, pulsed optomechanics~\cite{Vanner2011} provides a path to achieve such operations~\cite{ringbauer2018generation} and grow large spatial superpositions~\cite{clarke2018growing}. In this case, the superconducting qubit would be replaced by an optical mode and the qubit measurement statistics by photon counting statistics. Furthermore, as well as being a potential platform to test the collapse of quantum superpositions, the displacemon device may also offer a promising route towards tests of unconventional decoherence mechanisms in entangled systems. In Ref.~\onlinecite{kiesewetter2017pulsed}, it was proposed that Furry's hypothesis~\cite{furry1936note}---regarding spatially dependent decoherence mechanisms---could be tested in entangled optomechanical systems. Meanwhile, a method for generating two-mode entangled Schr\"{o}dinger-cat states via a measurement operator equivalent to a two-mode version of Eq.~\eqref{eq:gratingtransformation} was proposed in Ref.~\onlinecite{kanari2021non}, which suggests a two-mode displacemon system could be used to test Furry's hypothesis.

\section{Experimental considerations}
\label{sec:Experimental considerations}

The experiment proposed here integrates elements similar to those currently employed in hybrid nanomechanics, but operates in a largely unexplored parameter space.
Here we discuss briefly the main experimental challenges.
First there is the challenge of integrating comparatively large moving elements into a transmon qubit.
In three-dimensional cQED~\cite{Paik2011}, the transmon is typically $\sim 1$~mm long, so it would need to be made 1 or 2 orders of magnitude larger in order to accommodate resonators A and B.
There is no fundamental reason why this should not work, but it would require modifications to the geometry in order to keep the various capacitances involved within the desirable range for cQED.
Another geometrical challenge is that the thickness $D$ envisaged in Table~\ref{tab:parameters} is quite large compared to the normal thickness of the superconducting layer in a transmon, which is typically less than 100~nm.
Having fabricated the qubit, it will also be necessary to operate it coherently away from its flux sweet spot in order to achieve good electromechanical coupling.

The requirement to suppress thermal decoherence to the point that other mechanisms become observable means that the experiment must be performed at a very low temperature.
In a dilution refrigerator, if vibrational noise is sufficiently suppressed, a mechanical resonator typically~\cite{Regal2008} thermalises to $\Teff \sim 50$~mK.
With the aid of nuclear magnetic demagnetization, a 15~MHz resonator was recently~\cite{Cattiaux2021} cooled to $\Teff=~0.5$~mK.

Equally important with the cryostat temperature is the mechanical noise at the resonator frequency, which heats the resonator in the same way as an elevated environmental temperature.
The environmental noise can be quantified by a single-sided spectral density $S_{XX}(f)$, which describes the motion of the substrate to which the resonator is clamped.
Converting this position spectral density to a force spectral density $S_{FF}(f) = m\Omega^2 S_{XX}(f)$, and using equipartition to relate the resulting motion to an effective temperature leads to the relation~\cite{Poot2012}
\begin{equation}
    \Teff^\mathrm{mech} = \frac{1}{\kB} \frac{Q\Omega^3}{4} S_{XX}(2\pi\Omega)
\end{equation}
where $\Teff^\mathrm{mech}$ is the mechanical contribution to the effective environmental temperature $\Teff$.
In general the resonator will experience an environmental temperature that is determined by the larger of $\Teff^\mathrm{mech}$ and the cryostat temperature.
For example, the 8.7~mK effective temperature necessary to test gravitational collapse in Device A with $\sigma=\sigma_\mathrm{nuc}$ requires
\begin{equation}
    S_{XX}(2\pi \Omega) \lesssim \left(3.5 \times 10^{-16} \mathrm{m/\sqrt{Hz}}\right)^2,
\end{equation}
while for Device B, with $\Teff=1.5$~K, the requirement is
\begin{equation}
    S_{XX}(2\pi \Omega) \lesssim \left(1.6 \times 10^{-13} \mathrm{m/\sqrt{Hz}}\right)^2~.
\end{equation}
For comparison, the isolation platform built for Advanced LIGO has a vibration level below $10^{-12}~\mathrm{m/\sqrt{Hz}}$ at 8 Hz and below $10^{-13}~\mathrm{m/\sqrt{Hz}}$ at 100 Hz, the highest frequency measured~\cite{Matichard2015}. Thus, minimizing acoustic noise by mechanically isolating the sample and even the cryostat, and operating the resonator at an optimal frequency, will be important to testing collapse mechanisms in this frequency band.

\section{Conclusion}
The displacemon-based collapse testing protocol we introduce consists of a cooling stage followed by a grating operation on the resonator to prepare a non-Gaussian mechanical state of motion. The measurement on the mechanically-coupled superconducting qubit induces this grating operation and occurs over a timescale much less than the mechanical period. Therefore, our protocol is not constrained by the decoherence time of the qubit. The application of a second grating after an integer number of half-mechanical periods allows obtaining qubit measurement statistics that are sensitive to the underlying decoherence mechanisms. In principle, an experiment with the displacemon could be used to sense deviations from standard quantum mechanics predicted by any collapse model, regardless of whether a dynamical theory is proposed (as in GRW, CSL, and \Diosi's theory) or not (as in Penrose's theory). In this paper, we have explored the sensitivity of the displacemon device to the DP and CSL collapse models specifically. 

For the DP model, we have considered three possibilities for the gravitational radius, namely the nuclear radius, the zero-point amplitude of the beam resonator, and the zero-point amplitude of a single constituent aluminium atom. We find that by minimising their effective temperature, Devices A and B may be able to test for collapse in the first case (gravitational radius equals nuclear radius). Due to its lower frequency, Device B may also be suitable for testing the second case (gravitational radius equals zero-point amplitude of the beam). However, owing to the much larger length scale, our proposal is not capable of testing collapse when the gravitational radius equals the zero-point amplitude of a single aluminium atom within the lattice.

For the CSL model, we have taken into account the effect of the beam geometry on the momentum diffusion rate induced by CSL collapse. Moreover, we have derived an analytic result for the qubit measurement statistics in our testing protocol. We do so using solutions to the quantum Langevin equations for the system. This allows us to study the sensitivity of our testing protocol to the displacemon and CSL parameters, and identify regions of parameter space where CSL can be investigated.  While the parameters chosen for Device A only permit CSL to be tested at temperatures which are beyond current capabilities, the smaller mechanical frequency of Device B allows a larger range of $\lambda_{\mathrm{CSL}}$ to be explored in temperature regimes which could be reached with near-future devices. In tests of both DP and CSL collapse,  minimising the sensitivity of the devices to mechanical noise is likely to be the biggest challenge a future experiment will face.

We have shown that with technical advances in the fabrication of hybrid nanomechanical devices and minimisation of vibration sensitivities, the displacemon device provides an experimentally feasible route for testing collapse mechanisms in certain regions of parameter space. Finally, our testing protocol is applicable to other mechanical resonator devices, which can implement an equivalent grating operation and demonstrate sensitivity to decoherence in their measurement statistics. Thus, our proposal opens new avenues for tests of objective collapse models, complementary to matter-wave interferometery experiments.

\section*{Acknowledgements}
We are grateful to Kiran E. Khosla and Alistair W.~R. Smith for insightful discussions.
We acknowledge funding from the ERC (grant agreement 818751), the EU H2020 European Microkelvin Platform (grant agreement 824109), the Science and Technology Facilities Council (ST/T006102/1), the
Engineering and Physical Sciences Research Council (EP/T031271/1, EP/L016524/1), UK Research and Innovation (MR/S032924/1), and the Royal Society.

\section*{Author declaration}
The authors have no conflicts to disclose.

\section*{Data availability}
The data that supports the findings of this study are available within the article.

\vfill

\appendix
\section{CSL and thermal decoherence}
\label{app:CSL}
Here we outline the steps taken to obtain the probability given in Eq.~\eqref{eq:probability_decay}. Using Eq.~\eqref{eq:gratingtransformation}, the probability $P(k,\mathcal{D})=\braket{\Upsilon_{+}^\dag(t_k)\Upsilon_{+}(t_k)}$ can be written as:
\begin{equation}
\begin{split}
    P(k,\mathcal{D})&=\frac{1}{4}\braket{\cos^2\big(|\alpha|x(t_k)\big)}\\
    &=\frac{1}{2}+\frac{1}{2}\braket{\cos\big( 2|\alpha|x(t_k)\big)}
    \label{eq:prob_decay_first_step}
    \end{split}
\end{equation}
where the operator $x(t)=X(t)/X_{\mathrm{ZP}}$ is dimensionless. Therefore we can solve the quantum Langevin equations (written in dimensionless units) to obtain the following solutions evaluated at $t_k=k\pi/\omegam$:
\begin{subequations}
  \begin{align}
  \begin{split}
      x(t_k)&=\mathrm{e}^{-\gamma t/2} x(0)+\mathrm{e}^{-\gamma t/2}\Delta x(t_k)~,
      \end{split}
      \label{eq:Langevin_solution_X}\\
      \begin{split}
      p(t_k)&=\mathrm{e}^{-\gamma t/2}p(0)+\mathrm{e}^{-\gamma t/2}\Delta p(t_k)~,
      \end{split}
      \label{eq:Langevin_solution_P}
  \end{align}
\end{subequations}
where
\begin{subequations}
\begin{align}
    \Delta x(t_k)&=\int_0^{t_k} dt' \mathrm{e}^{\gamma t'/2}\sin(\omegam(t_k-t'))\xi(t')~,\label{eq:Delta_X}\\
  \begin{split}
  \Delta p(t_k)&=\int_0^{t_k} \mathrm{e}^{\gamma t'/2}\xi(t')\\&~~~~\times[\cos(\omegam(t_k-t'))-\frac{\gamma}{2\omegam} \sin(\omegam (t_k-t'))] \label{eq:Delta_P}
  \end{split}
\end{align}
\end{subequations}
and $\xi(t)=\sqrt{2}X_{\mathrm{ZP}}f(t)/\hbar$ such that $f(t)=f_{\mathrm{th}}(t)+f_{\mathrm{CSL}}(t)$. Substituting in these solutions into Eq.~\eqref{eq:prob_decay_first_step} and making use of the Isserlis-Wick theorem: 
\begin{equation}
\begin{split}
    \braket{\xi(t_1) \xi(t_2)\ldots \xi(t_n)}&=\frac{(2s)!}{2^s s!}\braket{\xi (t) \xi(t')}\\
    &=\frac{(2s)!}{2^s s!}(\sqrt{2}X_{\mathrm{ZP}})^2(D_{\mathrm{th}}+D_{\mathrm{CSL}})~,
    \end{split}
\end{equation}
(for $n=2s$, otherwise the expression evaluates to 0) we arrive at the solution: 
\begin{equation}
P(k,\mathcal{D})=\frac{1}{2}+\frac{1}{2}P_0(t_k)\exp\bigg[-8\mathcal{D}|\alpha|^2\frac{(1-e^{-k\pi\gamma/\omegam})}{\gamma(4+\gamma/\omegam)}\bigg]
\label{eq:prob_full_app}
\end{equation}
with
\begin{equation}
\begin{split}
    P_0(0)&=\frac{1}{2(1+e^{-2(1+2\bar{n})|\alpha|^2})}\times\\
    &~~~~\bigg[e^{-2(1+2\bar{n})|\alpha|^2(\kappa^2+2\kappa+1)}\\
    &~~~~+e^{-2(1+2\bar{n})|\alpha|^2(\kappa^2-2\kappa+1)}+e^{-2(1+2\bar{n})|\alpha|^2\kappa^2}\bigg]
    \end{split}
    \label{eq:P0}
\end{equation}
and $\kappa=e^{k\pi\gamma/(2\omegam)}$. In the limit of a high quality factor $Q=\Omega/\gamma\gg 1$ we can expand the term in the exponential in Eq.~\eqref{eq:prob_full}:
\begin{equation}
    -8\mathcal{D}|\alpha|^2 \frac{(1-e^{-k\pi/Q})}{\gamma(4+1/Q)}\approx -2\mathcal{D}\frac{k\pi}{\omegam}
\end{equation}
and thus obtain Eq.~\eqref{eq:probability_decay}.

\section*{References}

\bibliography{Bibliography}	

\end{document}